\documentclass[12pt,preprint]{aastex}
\usepackage{amsmath,natbib,psfig,emulateapj5}

\input{buote.defs}

\begin{document} 

\title{XMM-Newton and Chandra Observations of the Galaxy Group NGC
5044.\\ I. 
Evidence for Limited Multi-Phase Hot Gas}
\author{David A. Buote\altaffilmark{1}, Aaron
D. Lewis\altaffilmark{1}, Fabrizio Brighenti\altaffilmark{2,3}, \&
William G. Mathews\altaffilmark{2}} 
\altaffiltext{1}{Department of Physics and Astronomy, University of California
at Irvine, 4129 Frederick Reines Hall,\\ Irvine, CA 92697-4575}
\altaffiltext{2}{UCO/Lick Observatory, Board of Studies in Astronomy
and Astrophysics, University of California, Santa Cruz, CA 95064}
\altaffiltext{3}{Dipartimento di Astronomia, Universit\`a di Bologna,
via Ranzani 1, Bologna 40127, Italy}

\slugcomment{To Appear in The Astrophysical Journal}

\begin{abstract}

Using new \xmm\ and \chandra\ observations we present an analysis of
the temperature structure of the hot gas within a radius of 100~kpc of
the bright nearby galaxy group NGC 5044.  A spectral deprojection
analysis of data extracted from circular annuli reveals that a
two-temperature model (2T) of the hot gas is favored over single-phase
or cooling flow ($\mdot=4.5\pm 0.2$~\msunyr) models within the central
$\sim 30$~kpc.  Alternatively, the data can be fit equally well if the
temperature within each spherical shell varies continuously from $\sim
\thot$ to $\tcool \sim \thot/2$, but no lower.  The high spatial
resolution of the \chandra\ data allows us to determine that the
temperature excursion $\thot \rightarrow \tcool$ required in each
shell exceeds the temperature range between the boundaries of the same
shell in the best-fitting single-phase model.  This is strong evidence
for a multiphase gas having a limited temperature range. We do not
find any evidence that azimuthal temperature variations within each
annulus on the sky can account for the range in temperatures within
each shell. We provide a detailed investigation of the systematic
errors on the derived spectral models considering the effects of
calibration, plasma codes, bandwidth, variable \nh, and background
rate. We find that the RGS gratings and the EPIC and ACIS CCDs give
fully consistent results when the same models are fitted over the same
energy ranges for each instrument. The cooler component of the 2T
model has a temperature ($\tcool \sim 0.7$ keV) similar to the kinetic
temperature of the stars.  The hot phase has a temperature ($\thot
\sim 1.4$ keV) characteristic of the virial temperature of the $\sim
10^{13}$ $M_{\odot}$ halo expected in the NGC 5044 group.  However, in
view of the morphological disturbances and X-ray holes visible in the
\chandra\ image within $R \approx 10$ kpc, bubbles of gas heated to
$\sim \thot$ in this region may be formed by intermittent AGN
feedback.  Some additional heating at larger radii may be associated
with the evolution of the cold front near $R \sim 50$ kpc, as
suggested by the sharp edge in the EPIC images.

\end{abstract}


\section{Introduction}
\label{intro}

The hot gas in groups and clusters of galaxies is a vital window on
the history of star formation and metal enrichment in the
Universe. But to discover this history requires first a measurement of
the thermodynamic properties of the hot gas in these systems.  The
temperature structure of the gas is of special importance since it is
required to determine quantities such as the entropy and the metal
abundances. Galaxy groups are especially well-suited for studies of
their temperature structure since their $\sim 1$~keV temperatures
occur near the peak in sensitivity of current X-ray CCDs and near the
strong, highly temperature-sensitive iron L-shell emission lines.
 
Using new \xmm\ and \chandra\ observations we present an analysis of
the temperature structure of the hot gas of NGC 5044, perhaps the
brightest group in soft X-rays. Although several previous
\rosat\ and \asca\ studies demonstrated that the hot gas within $r\sim
30$~kpc is not isothermal \citep[e.g.,][]{davi94,buot99a}, the data
could not distinguish between single-phase and multiphase models. The
combined spatial and spectral resolution of the \xmm\ and
\chandra\ CCDs allows for unprecedented mapping of the temperatures
and elemental abundances of the hot gas in groups and clusters.  The
higher energy resolution, sensitivity, and larger field-of-view of the
\xmm\ EPIC CCDs are better suited for constraining the spatial and
spectral properties of the diffuse hot gas out to radii well past the
optical extent of the central galaxy; i.e., out to $r\sim 100$~kpc
assuming a distance of 33~Mpc using the results of \citet{tonr01} for
$H_0=70$~\kmsmpc\ (note: $1\arcsec=0.160$~kpc). The $\sim 1\arcsec$
resolution of \chandra\ is particularly useful for addressing the
properties of the hot gas on smaller scales.

In this paper we analyze the temperature structure of the hot gas in
NGC 5044. The metal abundances \citep[][hereafter Paper 2]{buot03b}
and gravitating mass distribution are discussed in companion papers.

\section{Observations and Data Preparation}
\label{obs}

\subsection{XMM}

NGC 5044 was observed with the EPIC pn and MOS CCD cameras for
approximately 20~ks and 22~ks respectively during AO-1 as part of the
\xmm\ Guest Observer program.  We generated calibrated events lists
for the data using the standard SAS v5.3.3 software. Since the diffuse
emission of NGC 5044 fills the entire field of view we estimate the
background using the standard ``background templates''. These
templates are events lists obtained by combining several high Galactic
latitude pointings\footnote{See \xmm\ calibration note by D. Lumb
(XMM-SOC-CAL-TN-0016).}.

Inspection of the CCD light curves for events with energies above
10~keV does not reveal any strong flares in the NGC 5044 observation,
but there are times of increased activity which are more pronounced in
the pn data.  After applying the count-rate screening criteria
recommended to match the background templates (and also the standard
screening criteria recommended for all observations), we arrive at
final exposures of 19.5~ks for the MOS1, 19.3~ks for the MOS2, and
8.9~ks for the pn.

Since the quiescent background varies typically by $\sim 10\%$ it is
necessary to normalize the background templates to each source
observation. Since NGC 5044 has a gas temperature $\sim 1$ keV there
is little emission from hot gas for energies $>5$~keV. We renormalized
the MOS1 and MOS2 background templates by comparing source and
background counts in the 7-12 keV band extracted from regions near the
edges of the MOS fields. We obtain background normalizations that are
\%16 and \%5 above nominal respectively for the MOS1 and MOS2.

Since the strict events screening mentioned above leaves the pn with
$<50\%$ of its raw exposure, it is evident that background flares
contaminate the pn more than the MOS. Even after the strict screening,
we still require a background normalization that is 21\% above nominal
for the pn.  Although this fraction is not much larger than for the
MOS1, an excess of the source over the background is clearly visible
above energies of a few keV. We have therefore also investigated a
potentially more accurate method of subtracting the flaring background
following our study of the \xmm\ observation of NGC 1399
\citep{buot03d}. That is, after subtracting the pn spectra taken from
regions near the edge of the field with the standard background
templates re-scaled by their nominal exposures, we fitted the
resultant spectrum with a two-component model consisting of a thermal
component, represented by an \apec\ thermal plasma, and a broken
power-law (BPL) model, representing the residual flaring background
which is most pronounced at high energies. For the BPL model we obtain
a break energy of 0.6~keV with power-law indices of -2.6 below the
break and 0.54 above the break; i.e., over most of the bandpass of
interest ($>0.6$~keV) the flaring background is well-described by a
power-law with index 0.54.

This BPL model defines the shape of the excess background above the
standard templates. We determined its normalization separately for
each region of interest using data between 6.0 and 7.25~keV. This
6~keV lower limit is selected to avoid contamination from softer
source emission while the upper limit is chosen to avoid calibration
emission lines.  We find that this method to subtract the excess
background in the pn data gives results consistent with the simpler
method of renormalizing the background template. In this paper we use
the BPL model of the excess background as our default method for the
pn. 

\subsection{Chandra}

NGC 5044 was observed by \chandra\ with the ACIS-S3 camera for
$\approx 22$~ks during AO-1. The events list was corrected for
charge-transfer inefficiency according to \citet{town02}, and only
events characterized by the standard \asca\
grades\footnote{http://cxc.harvard.edu/udocs/docs/docs.html} were
used. The standard \ciao\footnote{http://cxc.harvard.edu/ciao/}
software (version 2.2.1) was used for most of the subsequent data
preparation.

Since the diffuse X-ray emission of NGC 5044 fills the entire S3 chip,
we used the standard background
templates\footnote{http://cxc.harvard.edu/cal} to model the
background. After running the standard {\sc lc\_clean} script to clean
the source events list of flares with the same screening criteria as
the background templates, we arrive at a final exposure time of
$20.2$~ks for NGC 5044.

\section{Image and Radial Profile}

\label{image}

\begin{figure*}[t]
\parbox{0.49\textwidth}{
\centerline{\psfig{figure=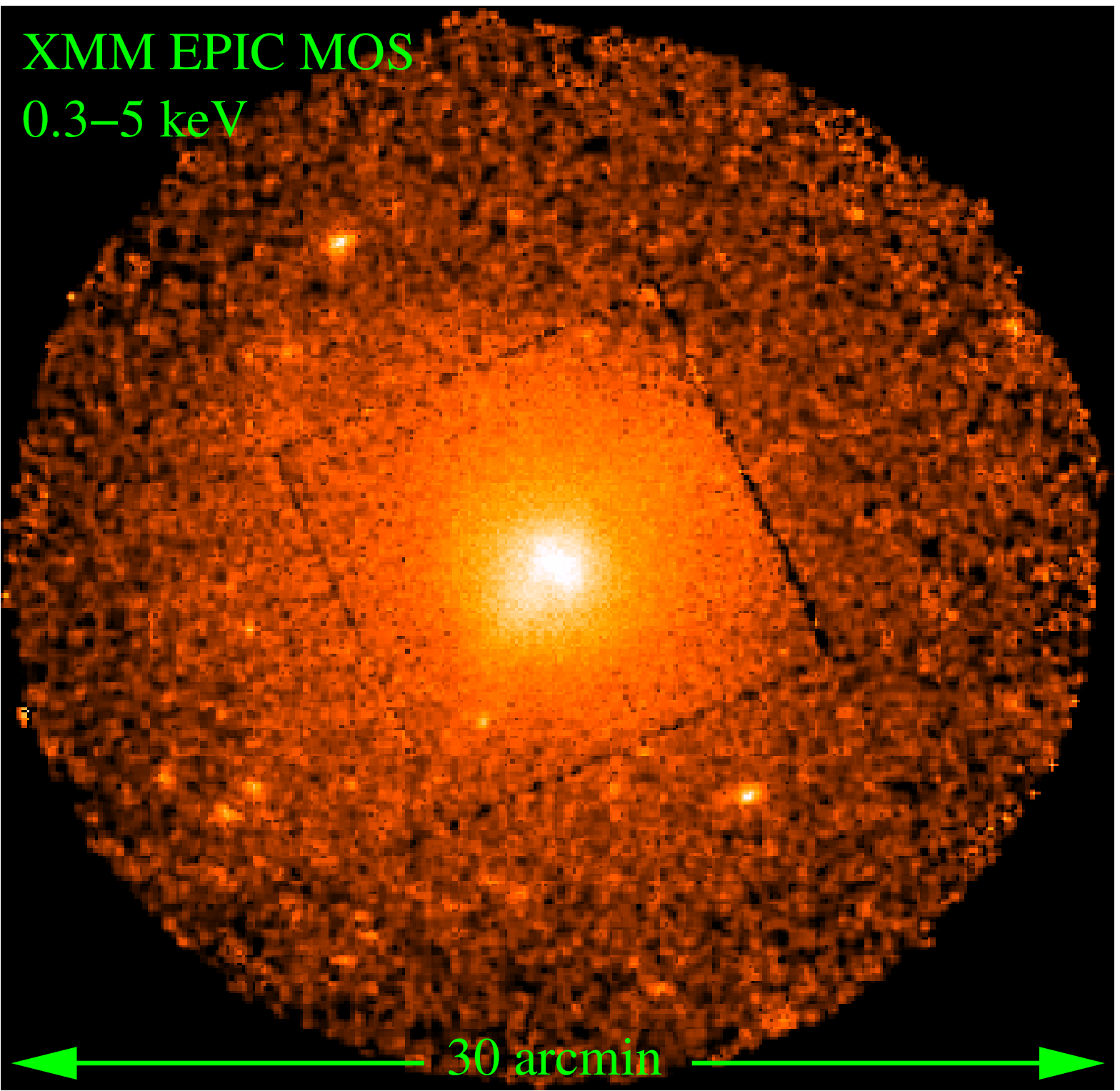,height=0.35\textheight}}
}
\parbox{0.49\textwidth}{
\centerline{\psfig{figure=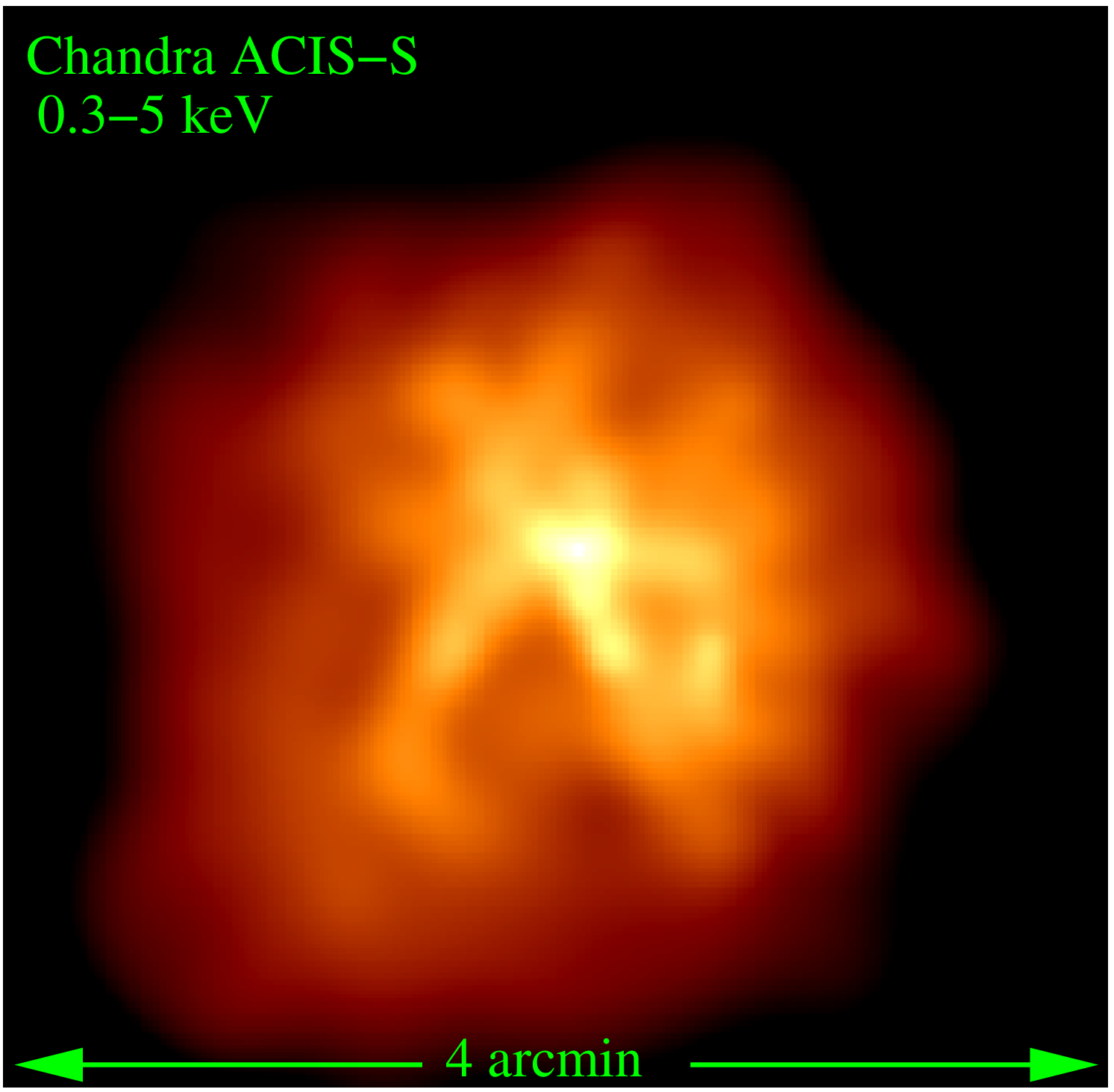,height=0.35\textheight}}
}
\caption{\label{fig.images} ({\sl Left panel}) False-color mosaic of
the \xmm\ MOS1 and MOS2 images of NGC 5044 adaptively smoothed using
the SAS task {\sc asmooth}. The image has been divided by the summed
MOS1 and MOS2 exposure maps to correct for exposure variations. ({\sl
Right panel}) Portion of the archival \chandra\ ACIS-S image (also
false-color) located at the center of NGC 5044 adaptively smoothed
using the CIAO task {\sc csmooth}. In both images celestial N is up
and E is to the left.} \end{figure*}

In Figure \ref{fig.images} we display an adaptively smoothed mosaic of
the \xmm\ MOS1 and MOS2 images as well as the adaptively smoothed
\chandra\ ACIS-S3 image. Several sources can be seen embedded
in the diffuse emission which fills the entire MOS field. That we see
more sources at large radius suggests that discrete sources at smaller
radii are swamped by the diffuse emission which is brightest near the
center. A careful inspection of the MOS and pn images within $R\approx
1\arcmin$ of NGC 5044 reveals significant non-circular features. The
higher-resolution \chandra\ ACIS-S image (Figure \ref{fig.images})
shows these features to be holes in the hot gas similar to those seen
in \chandra\ images of the cores of galaxy clusters
\citep[e.g., ][]{fabi00_perseus,mcna00,chur01}. Typically these holes
are filled by extended radio emission that emanates from a central
source. However, the ($\approx 45\arcsec$) NVSS \citep{cond98} radio
image of NGC 5044 reveals only a fairly bright central point-like
source suggestive of an AGN. Perhaps a deeper observation at higher
resolution would find extended radio emission filling the X-ray holes.

Over a radius $R\approx 1-5\arcmin$ (i.e., to the edge of the central
MOS CCD) the diffuse emission is approximately circular with
concentric isophotes. Between $R\approx 5-6\arcmin$ the isophote
centroid is slightly offset from values at smaller radius. The origin
of this shift is apparent in Figure \ref{fig.images} as excess
emission to the East and N-E outside the central CCD.  This offset was
noted previously in the \rosat\ PSPC observation of NGC 5044
\citep{davi94}. The \xmm\ image, however, also reveals that there is a
sharp edge in the X-ray surface brightness on the West, N-W side near
this offset region, $R\sim 6\arcmin$. This type of sharp feature has
been observed recently in \chandra\ observations of several galaxy
clusters and has been interpreted as a ``cold front''; i.e., a contact
discontinuity between regions of different density and temperature
\citep[see][]{mark02a}. We note that at radii larger than $\sim
6\arcmin$ the isophote centroids are consistent with those for $R \la
5\arcmin$.

\begin{figure*}[t]
\parbox{0.49\textwidth}{
\centerline{\psfig{figure=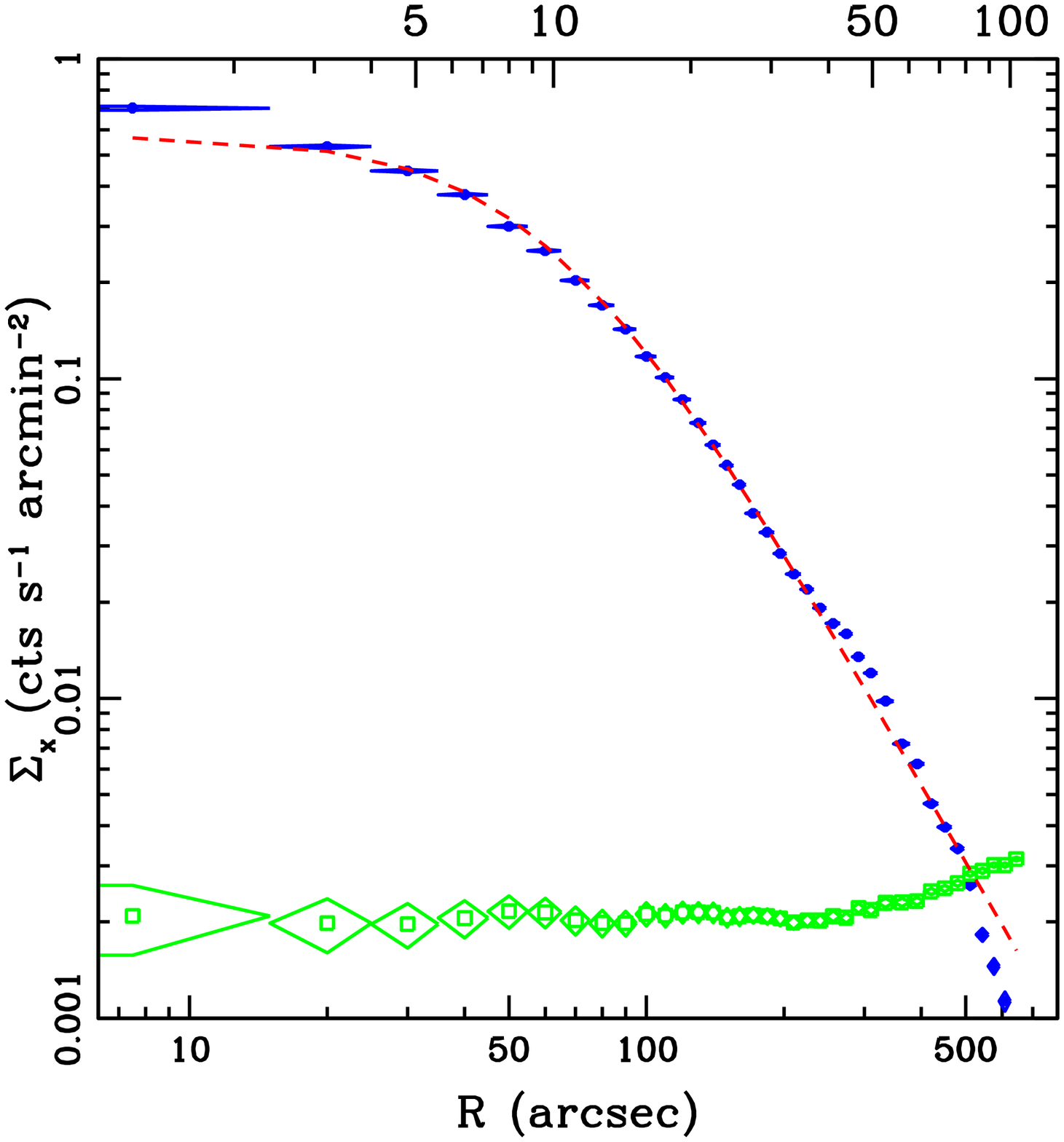,height=0.35\textheight}}
}
\parbox{0.49\textwidth}{
\centerline{\psfig{figure=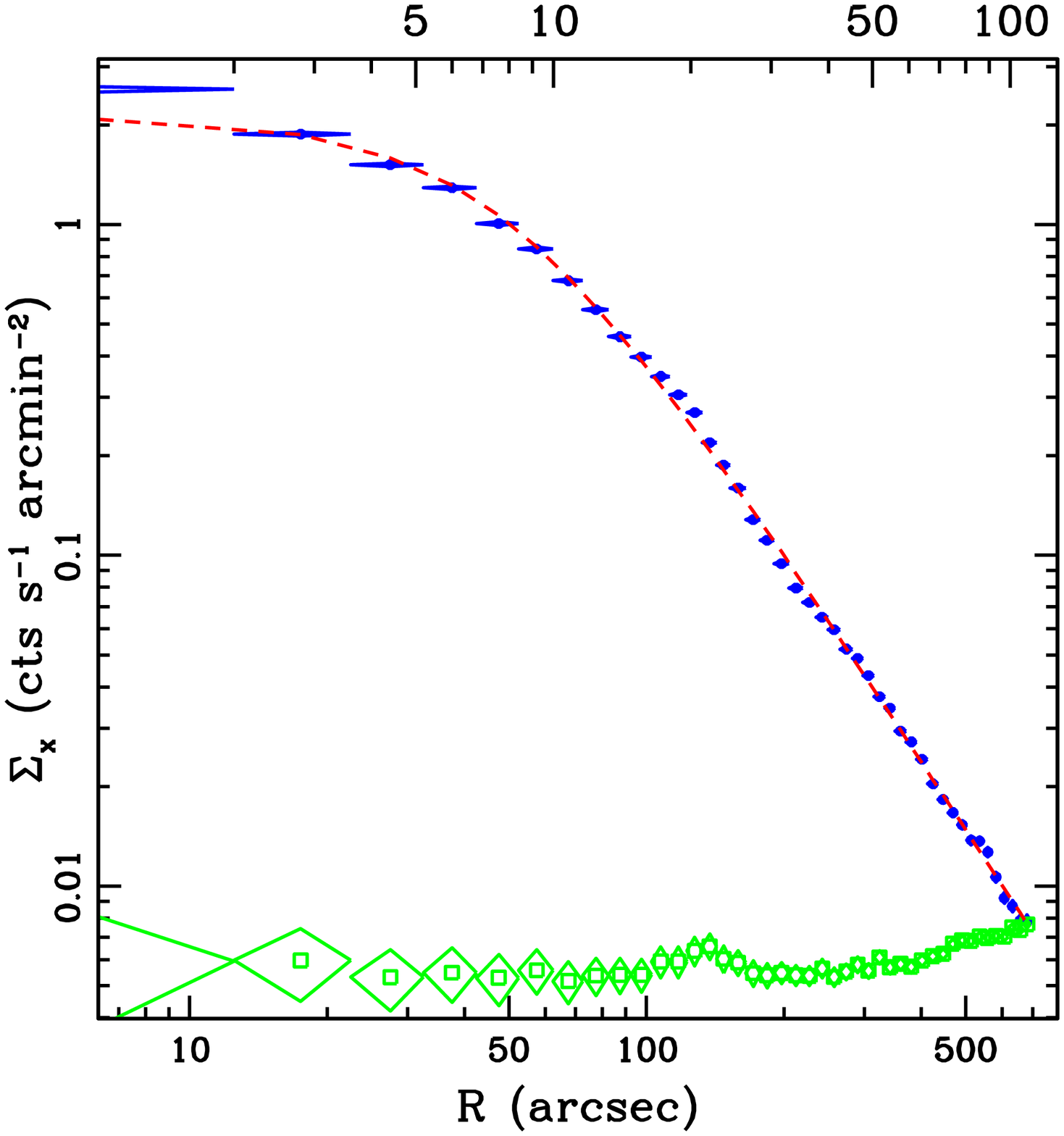,height=0.35\textheight}}
}
\caption{\label{fig.radpro} ({\sl Left panel}) Azimuthally averaged
0.3-5~keV surface brightness profile of the background-subtracted,
exposure-corrected combined MOS1 and MOS2 data (kpc units on top
axis). The lower points represent the combined background values. The
best-fitting $\beta$ model for the MOS data is indicated by the dashed
line. ({\sl Right panel}) Same quantities are plotted for the pn
data.}
\end{figure*}

We characterized the image further by computing the azimuthally
averaged surface brightness profile in the 0.3-5~keV band separately
for the MOS and pn data. We created exposure maps for each detector
using the standard SAS software and used these maps to flatten the
images. We combined the MOS1 and MOS2 data into a single
exposure-corrected image.  The same procedures were followed for the
background images. (Since here we are not concerned with the detailed
spectral dependence of the background for the pn we used the
renormalized template accounting for the factor of 21\% discussed in
\S \ref{obs}.) Point sources and chip gaps were masked out from the
calculation.

The radial profiles, \radpro, of the source and background images are
shown in Figure \ref{fig.radpro}.  Within $R\sim 7\arcmin$ the radial
profiles are similar for both detectors, but at larger radii the
profiles appear to deviate. To facilitate comparison between the
profiles, we fitted standard ``$\beta$ models'' convolved with the
\xmm\ PSF\footnote{See XMM calibration note EPIC-MCT-TN-012 by
Ghizzardi.} to \radpro. The best-fitting models are shown in Figure
\ref{fig.radpro}. We obtained parameters, $r_{\rm c}=61.9\pm
0.4\arcsec$, and $\beta=0.585\pm 0.001$ for the MOS and, $r_{\rm
c}=47.9\pm 0.4\arcsec$, and $\beta=0.522\pm 0.001$ for the pn. The
values of these parameters confirm the impression from the figure that
at large radius \radpro\ is steeper for the MOS.

The steeper profile of the MOS at larger radii is probably the result
of small errors in the background estimate at large radius. Near
$R=7\arcmin$ the MOS background rate equals the source rate indicating
that the derived \radpro\ for the source is very sensitive to small
errors in the background for $R\ga 7\arcmin$. In contrast, the
background rate does not equal the source rate for the pn until near
the edge of the field, $R\sim 11\arcmin$. The robustness of \radpro\
for the pn is also supported by the analysis of the \rosat\ data of
NGC 5044 by \citet{davi95} who obtained $\beta=0.53\pm 0.02$
consistent with our result for the pn. If data with $R> 400\arcsec$
are excluded for the MOS, then we obtain best-fitting parameters,
$r_{\rm c}=52.1\arcsec$ and $\beta=0.542$, in better agreement with
the pn and \rosat\ values.

We conclude that a single $\beta$ model is a good average description
of the radial surface brightness profile of NGC 5044 between $R\sim
10\arcsec$ and $\sim 11\arcmin$. (At the center there is evidence for
a small excess above the $\beta$ model on the scale of the PSF. This
is also the region where the azimuthal distortions are most pronounced
in the \chandra\ image.)  Therefore, the possible ``cold front''
suggested by the isophotal centroid offset near $R=6\arcmin$ does not
substantially distort the radial profile away from that of an
equilibrium configuration.

\section{Spectral Deprojection Analysis}
\label{deproj}

Motivated by the regular appearance of the radial profile and the lack
of substantial azimuthal variations in the spectral properties, we
focus our analysis on the azimuthally (and spherically) averaged
spectral properties of the X-ray emission. We present an analysis of
the azimuthal spectral properties of the \chandra\ and \xmm\ data in
\S \ref{2d}.

\subsection{Preliminaries}
\label{prelim}

For each \xmm\ detector (MOS1,MOS2,pn) we extracted spectra in
concentric circular annuli located at the X-ray centroid of the inner
contours (i.e., computed within a $2\arcmin$ radius) such that the
width of each annulus contained $\approx 8000$ background-subtracted
counts in the 0.3-5 keV band in each MOS detector, and the minimum
width was set to $1\arcmin$ for PSF considerations; note that we
address residual effects of the PSF by comparison to fits of the
high-resolution \chandra\ data alone in \S \ref{chan} and \S
\ref{sys}. Obvious point sources were masked out before the extraction
(and corresponding regions also were omitted from the background
templates).  These restrictions resulted in a total of eight annuli
within $r=10\arcmin$; these annuli are defined in Table
\ref{tab.fits}.  (Note that the results we present below are quite
insensitive to the choice of radius within which to compute the
centroid.) RMF and ARF files were generated using SAS. (Note that the
vignetting correction is administered through the ARF file.) Finally,
the spectral pulse-invariant (PI) files for each annulus were rebinned
such that each energy bin contains a minimum of 30 counts appropriate
for $\chi^2$ fitting.

We first consider the \chandra\ data extracted in the same regions as
the \xmm\ data subject to the constraint that the regions lie entirely
within the ACIS-S3 field. This restriction gives \chandra\ data within
the first three radial bins defined for the \xmm\ data. Later in \S
\ref{chan} we shall examine the \chandra\ data within smaller spatial
regions.  We have applied the latest corrections to the \chandra\ ARF
files that account for a time-dependent degradation in the quantum
efficiency at lower energies (i.e., \ciao\ {\sc corrarf} routine).

To obtain the three-dimensional properties of the X-ray emitting gas
we perform a spectral deprojection analysis assuming spherical
symmetry using the (non-parametric) ``onion-peeling'' technique.  That
is, one begins by determining the spectral model in the bounding
annulus and then works inward by subtracting off the spectral
contributions from the outer annuli.  We have previously developed a
code based on \xspec\ \citep{xspec} to implement this procedure
\citep{buot00c}. Consequently, we obtain temperatures, abundances,
densities, and absorption column densities (and any other desired
parameters) by fitting spectral models to the ``deprojected spectra''.

In this procedure we account for the emission projected from shells
outside of our bounding annulus by assuming the X-ray emissivity for
such shells varies as a power-law with radius and has the same
spectral shape as determined from the bounding annulus. For the \xmm\
data for NGC 5044 we take the radial surface brightness to vary as
$r^{-2}$ (or $\beta=0.5$ for the $\beta$ model) as indicated by the
data within $R=10\arcmin$. As found previously (\S 3.1 of
\citealt{buot00c}) the derived spectral parameters (except density in
the penultimate annulus) are quite insensitive to this choice.

To estimate the uncertainties on the fitted parameters we simulated
spectra for each annulus using the best-fitting models and fit the
simulated spectra in exactly the same manner as done for the actual
data. From 20 Monte Carlo simulations we compute the standard
deviation for each free parameter which we quote as the ``$1\sigma$''
error. (This is a slightly different manner of quoting the errors
compared to \S 3.2 of \citealt{buot00c}.)

Because deprojection always inflates the errors between points, which
is related to the error associated with the derivative of the
emissivity in an Abel inversion, we regularize (i.e., smooth) some of
the parameters (see \S 3.3 of \citealt{buot00c}). In actuality we
regularize {\it ex post facto} by restricting the value of a parameter
within certain bounds specified by a pre-determined radial variation
for the parameter. We find it necessary to regularize only the O and
Ne abundances so that the radial abundance variation of the
logarithmic derivative is between $\pm 2$. We emphasize that
regularization only applies to O and Ne for the deprojected (3D)
spectral analysis. No regularization is applied to any 2D model.

We take the solar abundances in \xspec\ (v11.2.0af) to be those given
by the \citet{grsa} table which use the correct new photospheric value
for iron which agrees also with the value obtained from solar-system
meteorites \citep[e.g.,][]{mcwi97}.

\subsection{Simultaneous Fitting of XMM-Chandra Data}
\label{xmmchan}

\begin{table*}[t] \footnotesize
\caption{Quality of Spectral Fits ($\chi^2$/dof) for 1T and 2T Models
\label{tab.fits}} 
\begin{center} \vskip -0.4cm
\begin{tabular}{ccccccc} \tableline\tableline\\[-7pt]
& $R_{\rm in}$ & $R_{\rm out}$ & \multicolumn{2}{c}{1T} &
\multicolumn{2}{c}{2T}\\ 
Bin & (arcmin) & (arcmin) & 2D & 3D & 2D & 3D\\ 
\tableline \\[-7pt]
1 & 0.0   &  0.5  &  779.8/451  &  737.9/451  & 646.7/449  & 644.0/449  \\  
2 & 0.5   &  1.5  & 1812.3/689  & 1368.1/689  & 974.2/687  & 949.8/687  \\  
3 & 1.5   &  2.5  & 1338.2/693  & 1189.0/693  & 855.3/691  & 836.0/691  \\  
4 & 2.5   &  3.5  &  730.6/482  &  666.8/482  & 610.3/480  & 537.3/480  \\ 
5 & 3.5   &  4.5  &  535.6/454  &  504.7/454  & 521.7/452  & 484.4/452  \\ 
6 & 4.5   &  5.7  &  547.7/469  &  548.3/469  & 516.2/467  & 514.4/467  \\ 
7 & 5.7   &  7.6  &  538.0/515  &  572.4/515  & 519.7/513  & 550.0/513  \\ 
8 & 7.6   & 10.1  &  674.6/561  &  674.6/561  & 658.7/559  & 658.7/559  \\ 
\tableline \\[-1.0cm]
\end{tabular}
\tablecomments{Bin refers to an annulus for 2D models and to a shell
for 3D models. Single-temperature (1T) models are described in \S \ref{1t}
and two-temperature (2T) models are described in  \S \ref{2t}.}
\end{center}
\end{table*}

Within the eight annuli defined in Table \ref{tab.fits} we consider
simultaneous fits to the MOS1, MOS2, and pn \xmm\ data within radial
bins 4-8 for which we do not have \chandra\ data. Within radial bins
1-3 we perform simultaneous fits to the \chandra\ and \xmm\
data. Comparison of the \chandra\ and \xmm\ data within their overlap
regions is discussed in \S \ref{sys}. We focus on fits performed over
the energy range 0.5-5~keV from consideration of the iron abundances
obtained from separate \chandra\ and \xmm\ fits as discussed in Paper
2. However, all results presented for the temperatures in this paper
are consistent with those obtained when fitting 0.3-5~keV (\S
\ref{sys}).

Our baseline model (1T) consists of a single thermal plasma component
using the \apec\ code modified by foreground Galactic absorption
($\nh=5\times 10^{20}$~\cmsq) using the \phabs\ model in \xspec. The
free parameters for this baseline model are all associated with the
plasma component: temperature ($T$), normalization, and Fe, O, Ne, Mg,
Si, and S abundances -- all other elements tied to Fe in their solar
ratios.  (Models with intrinsic absorption are discussed in \S
\ref{sys}.) We obtained results for the 1T model fitted directly to the
data projected on the sky (i.e., traditional 2D model) and also fitted
to the deprojected data (i.e., 3D model) as discussed in \S
\ref{deproj}.

To account for small calibration differences between detectors we
multiplied the spectrum of each detector by a detector-dependent
constant. These constants were determined by fitting the 1T model to
the data with the constants as free parameters. We fixed the constants
to these values for all subsequent spectral fits. This approach
guarantees that the relative normalizations of multiple emission
components (e.g., in 2T models) are the same for each detector.

\subsubsection{Single-Temperature Models}
\label{1t}

\begin{figure*}[t]
\parbox{0.49\textwidth}{
\centerline{\psfig{figure=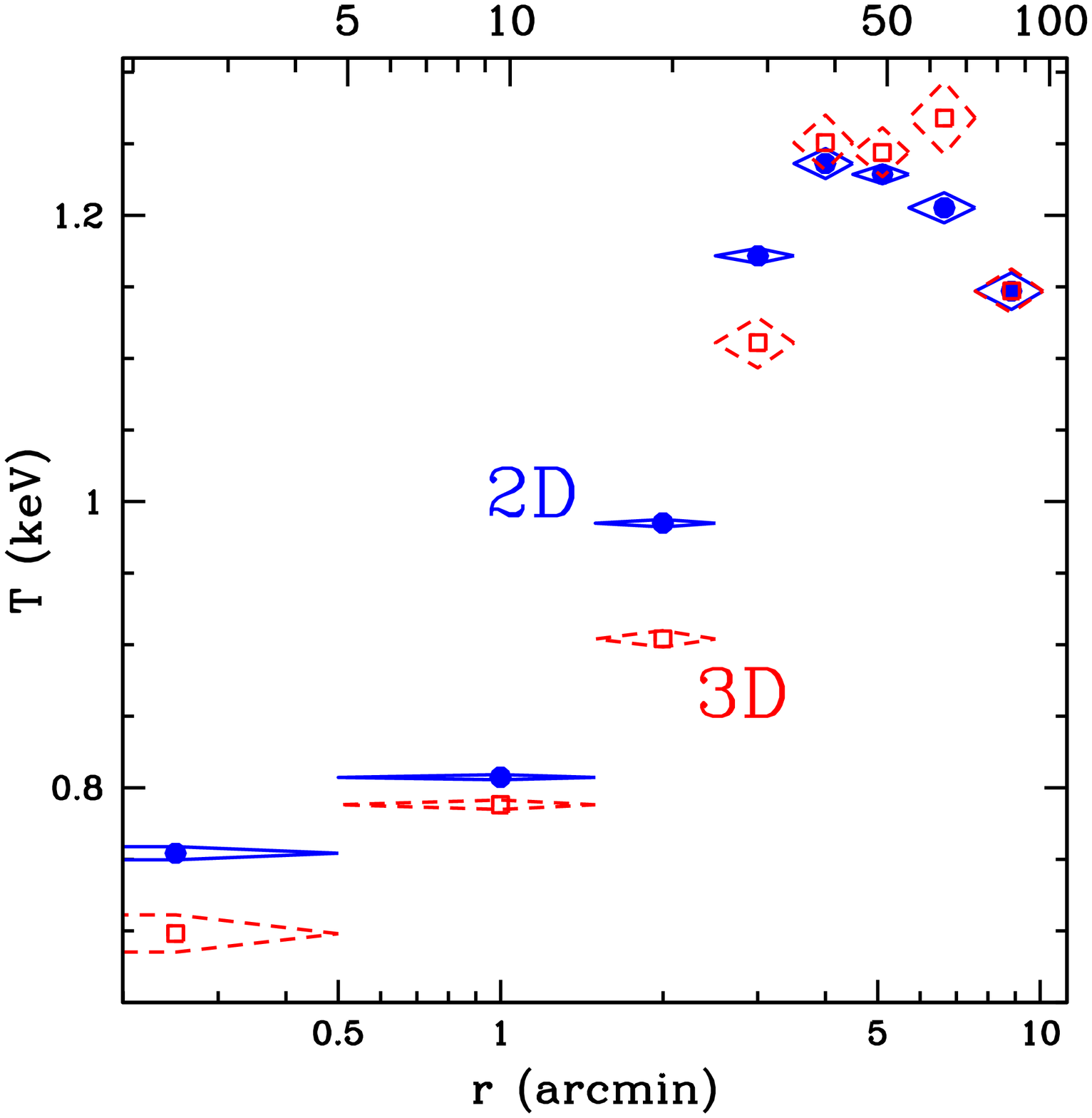,height=0.3\textheight}}
}
\parbox{0.49\textwidth}{
\centerline{\psfig{figure=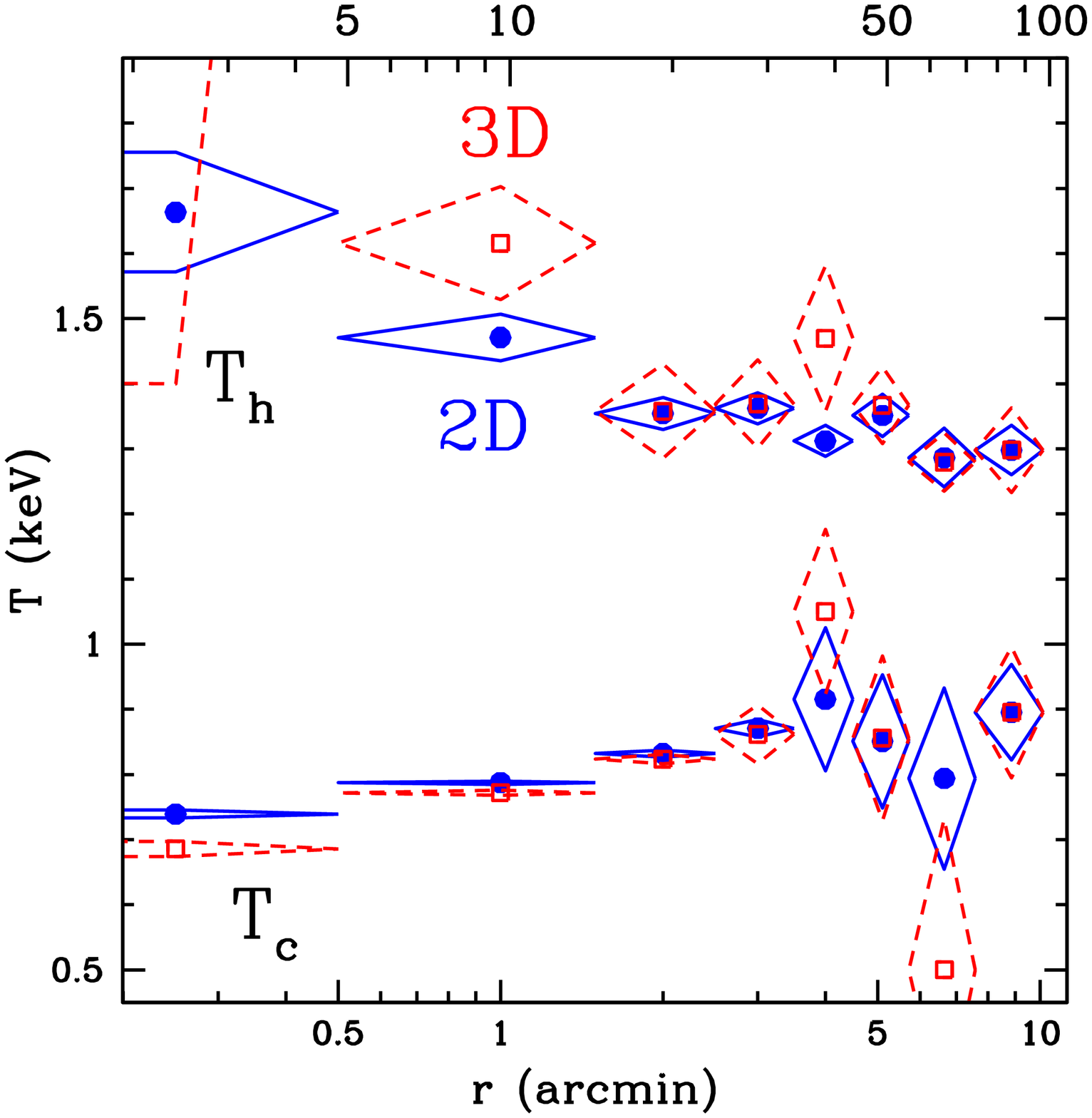,height=0.3\textheight}}
}
\caption{\label{fig.temp} Radial temperature profiles
(units -- bottom: arcminutes, top: kpc) and $1\sigma$ errors for ({\sl
Left panel}) 1T and ({\sl Right panel}) 2T models fitted
simultaneously to \xmm\ and \chandra\ data. Note that the \chandra\
data apply only to the inner three radial bins. In each case ``3D''
refers to results obtained from a spectral deprojection analysis. }
\end{figure*}

\begin{figure*}[t]
\centerline{\psfig{figure=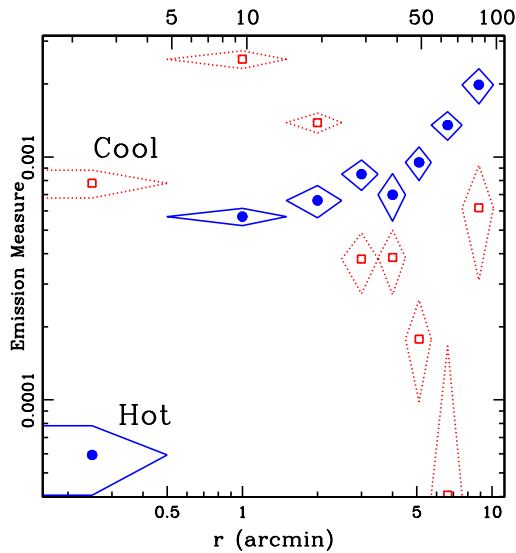,height=0.3\textheight}}
\caption{\label{fig.em} Emission measure of
the cooler and hotter components for the 2T (3D) model whose
temperatures are plotted in Figure \ref{fig.temp}.  The emission
measure is expressed as the normalization of the \apec\ plasma model
as given in \xspec; i.e., normalization is $10^{-14}n_en_pV/(4\pi
D_A^2(1+z)^2)$, where $n_e$ is the electron number density, $n_p$ is
the proton number density, $V$ is the emitting volume, $D_A$ is the
angular diameter distance, and $z$ is the redshift. The units are
arcminutes on the bottom axis and kpc on the top axis.}
\end{figure*} 

The temperature profiles for the 1T models (2D and 3D) are displayed
in Figure \ref{fig.temp}. For the 1T (2D) model the temperature rises
from $T\sim 0.7$~keV at the center to $T\sim 1.2$~keV at large radii
consistent with previous 2D \rosat\ determinations
\citep[e.g.,][]{davi94,buot00c}. The 3D temperature has a steeper rise
outward from the center consistent with the 2D profile being smeared
by projection effects. The lower temperature in the bounding annulus
certainly suggests a declining temperature value, but its value is
underestimated because of projection; i.e., we have had to assume the
projected emission from shells exterior to our bounding annulus has
the same spectral shape as the bounding annulus (\S
\ref{deproj}).

The $\chi^2$ values for both models are listed in Table
\ref{tab.fits}. This model is a reasonably good fit in the outer
annuli, but the fit degrades at progressively smaller radii, though
the trend reverses somewhat for the central annulus.  For most radial
bins the 2D and 3D fits are similar, with the 3D model usually giving
slightly better $\chi^2$ values. In shells 2-4, however, the 3D model
clearly provides a better fit indicating that annuli 2-4 contain
multiple temperature components projected on to the sky.

The 1T (3D) model is still a relatively poor fit in shells 1-4. In
Figure \ref{fig.spec.1t} we show the EPIC and ACIS spectra for annulus
2 fitted to the 1T (3D) model fits and residuals.  The 1T (3D) model
yields fit residuals near 1~keV that are fully characteristic of those
obtained when trying to force a single-temperature model to fit a
spectrum consisting of a multiple components with temperatures near
1~keV \citep[e.g.,][]{buot98c,buot99a,buot00a}. The small deviations
above 2~keV also suggest the presence of another (higher) temperature
component.

\subsubsection{Two-Temperature Models}
\label{2t}

Since the fits within the central regions -- even when the spectral
data are deprojected -- indicate the presence of at least one more
spectral component, we investigated fitting simple two-temperature
models (2T) to the data. The abundances for each component are tied
together to limit the number of free parameters and because relaxing
this condition does not amount to a substantial improvement in the fit
quality. (In Paper 2 we briefly discuss 2T models with separately
varying abundances on each component.) Consequently, the 2T models add
only two free parameters: the temperature and normalization of the
second component.

\begin{figure*}[t]
\parbox{0.49\textwidth}{
\centerline{\psfig{figure=f5a.eps,angle=-90,height=0.24\textheight}}
}
\parbox{0.49\textwidth}{
\centerline{\psfig{figure=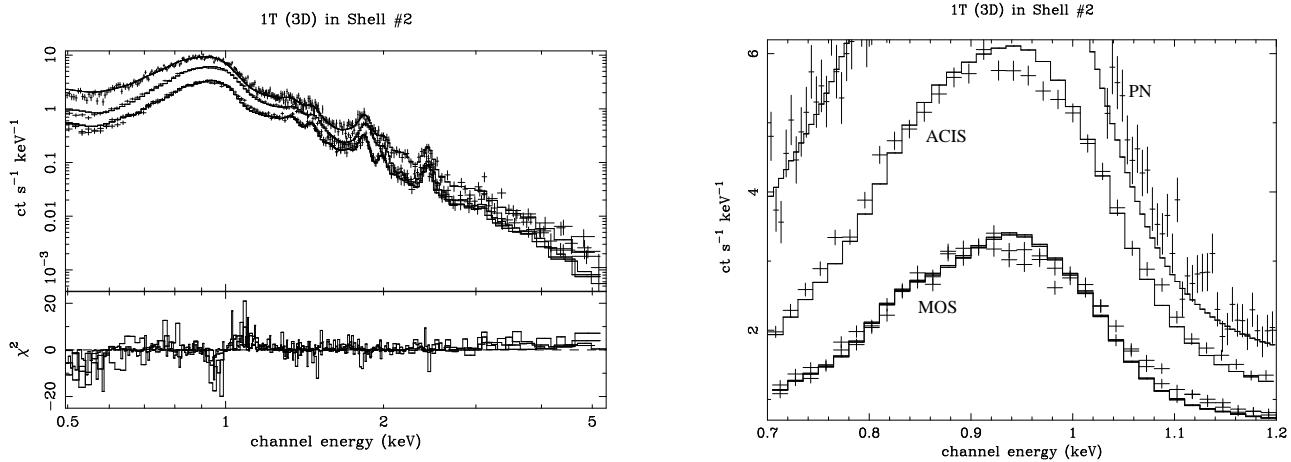,angle=0,height=0.25\textheight}}
}
\caption{\label{fig.spec.1t} ({\sl Left panel})  EPIC MOS1, MOS2, pn,
and ACIS-S3 spectra for annulus \#2 fitted with a 1T (3D) model. ({\sl
Right panel}) A close-up of the plot in the left panel emphasizing the
the MOS and ACIS data and model in the region of strong iron L-shell
lines.}
\end{figure*}

\begin{figure*}[t]
\parbox{0.49\textwidth}{
\centerline{\psfig{figure=f6a.eps,angle=-90,height=0.24\textheight}}
}
\parbox{0.49\textwidth}{
\centerline{\psfig{figure=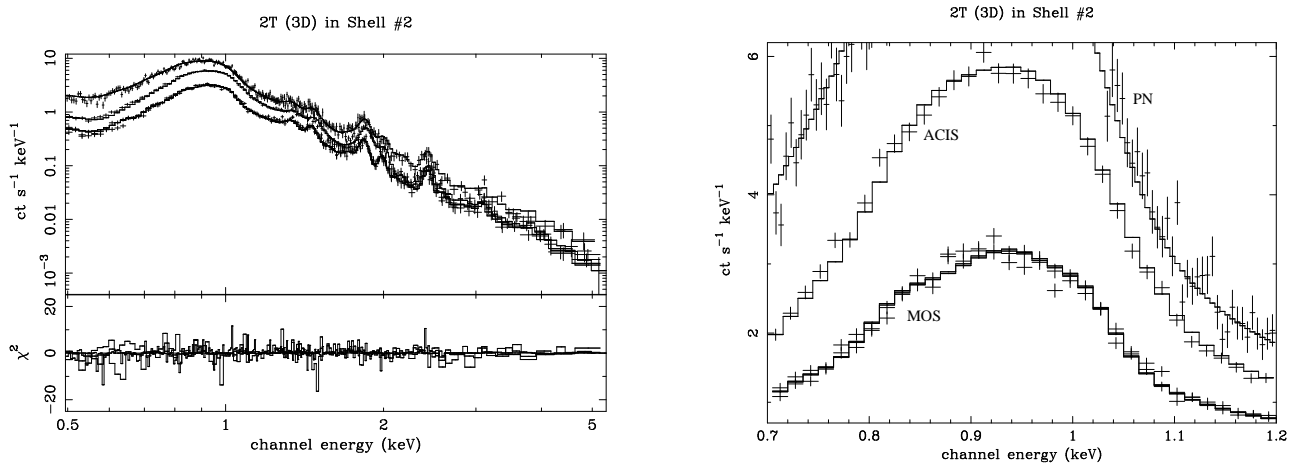,angle=0,height=0.25\textheight}}
}
\caption{\label{fig.spec.2t} ({\sl Left panel}) Same as Figure
\ref{fig.spec.1t} except that the 2T (3D) model is shown.}
\end{figure*}

The $\chi^2$ values for both the 2T (2D) and 2T (3D) models are listed
in Table \ref{tab.fits}. In all spatial bins the fits are improved by
the 2T models, with the greatest improvement occurring in the central
bins (1-4). The considerable improvement for bin 2 is shown in Figure
\ref{fig.spec.2t} for the 2T (3D) model. In comparison to Figure
\ref{fig.spec.1t} the fit residuals are reduced substantially, and
overall the fit is as good as could be expected for such a simple
model.

The values of $\chi^2$ obtained for the 2T (2D) and 2T (3D) models are
very similar, but the 3D models generally have lower $\chi^2$
values. Only in bin 4 is the 3D model an obvious improvement over the
2D version. Since 2T models can mimic radial temperature variations
accurately, this agreement is not unexpected. Nevertheless, because of
the similarity between the 2D and 3D fits, projection effects are not
very important for the 2T models.

The temperature profiles of the 2D and 3D 2T models are displayed in
Figure \ref{fig.temp}. Focusing for the moment on the 2D model which
has better constrained values than the 3D model, it is seen that both
the cool ($\tcool$) and hot ($\thot$) temperature components are
consistent with an isothermal radial profile for $R\ga 20$~kpc. For
smaller radii, $\tcool$ remains nearly constant with an indication of
a slight downturn in the central two bins. The hotter component rises
slightly at the center.  The values of $\tcool$ at small radii and the
values of $\thot$ at large radii are similar to the 1T temperature
values.

Within the 1-2~$\sigma$ errors the 3D and 2D temperatures are
consistent in all bins, but there are some trends that deserve
attention. At large radii ($r\ga 40$~kpc) \tcool\ is not tightly
constrained particularly for the 3D model; the exception is the final
bin where the 2D and 3D fits are equivalent in our deprojection
procedure (apart from the inferred normalization). In bin 7 the cooler
component is barely detected in 3D; the ratio of normalizations for
the best-fitting cooler-to-hotter components is 0.12 for the 2D model
compared to only 0.03 for the 3D model.  Also in bin 7 the value of
\tcool\ is poorly constrained for the 3D model, and the best-fitting
value of 0.5~keV is located at the lower range we allowed for during
the fit.  We attribute the stronger detection of the 2T model in bin 8
compared to bin 7 to the limitation in the way we treat the
deprojection of the bounding annulus; i.e., our deprojection method
assumes the source flux at large radii has the same spectrum as the
bounding shell.

At small radius ($r\la 10$~kpc), the temperature of the hotter
component rises and becomes poorly constrained in bin 1 for the 3D
model, where the best-fitting value is$T= 3.7$~keV and the lowest
value from 100 Monte Carlo simulations is 1.4~keV which is shown in
Figure \ref{fig.temp}.  It is possible that the rise in \thot\ in bin
2, and especially bin 1, arises from the emission of unresolved
discrete sources similar to the trend observed for the related system,
NGC 1399 \citep{buot02a}. If we add a 10~keV bremsstrahlung (brem)
component to the 2T models the fits are not improved, but the fitted
value of \thot\ in bins 1-2 are consistent with those adjacent
bins. Moreover, we obtain a luminosity $\lx \approx 3\times
10^{40}$~\ergcms\ for the brem component. Since this value is
comparable to the (relatively uncertain) luminosity expected from
discrete sources in NGC 5044 using the results of \citet{osul01a}, we
conclude that the unresolved emission from discrete sources is a
viable candidate for the rise in \thot\ at small radius in the 2D
models (though \thot\ is still a separate component from the discrete
sources).

In Figure \ref{fig.em} we plot the emission measures for each
component of the 2T (3D) model. The cooler component dominates within
$\sim 10$~kpc while the hotter component dominates for $r\ga
50$~kpc. Over $\approx 15-40$~kpc the relative emission measures of
the components are within a factor of two of each other.

\subsubsection{Cooling Flow and Other Multitemperature Models}
\label{dem}

\begin{table*}[t] \footnotesize
\caption{Quality of Spectral Fits ($\chi^2$/dof) for other Multitemperature Models
\label{tab.fits.dem}} 
\begin{center} \vskip -0.4cm
\begin{tabular}{cccc} \tableline\tableline\\[-7pt]
Shell & Cooling Flow & GDEM & PLDEM\\ \tableline \\[-7pt]
1 &   729.6/450  &   695.5/450    & 665.0/449 \\  
2 &  1375.6/688  &  1091.4/688    & 973.3/687 \\  
3 &   982.1/692  &   869.7/692    & 847.2/691 \\  
4 &   591.2/481  &   546.5/481    & 546.9/480 \\ 
5 &   490.7/453  &   484.6/453    & 494.7/452 \\ 
6 &   526.0/468  &   518.9/468    & 535.8/467 \\ 
7 &   561.1/514  &   554.3/514    & 558.1/514 \\ 
8 &   659.7/560  &   658.3/560    & 660.6/560 \\ 
\tableline \\[-1.0cm]
\end{tabular}
\tablecomments{All models are 3D. See \S \ref{dem} for description of the models.}
\end{center}
\end{table*}

\begin{table*}[t] \footnotesize
\caption{Selected Parameters for other Multitemperature Models
\label{tab.dem}} 
\begin{center} \vskip -0.4cm
\begin{tabular}{cc|cc|rcc} \tableline\tableline\\[-7pt]
& Cooling Flow & \multicolumn{2}{c}{GDEM} &
\multicolumn{3}{c}{PLDEM} \\[+2pt]
& \mdot\ & $T_0$ & \sigmat\ &
\multicolumn{1}{c}{$\alpha$} & $T_{\rm min}$ & $T_{\rm max} - T_{\rm min}$\\
Shell & (\msun\ yr$^{-1}$) & (keV) & (keV) & & (keV) & (keV)\\ 
\tableline \\[-7pt]
1 & $0.00\pm 0.15$  & $0.72\pm 0.01$  & $0.10\pm 0.04$   & $-5.0\pm 2.3$  & $0.62\pm 0.14$   & $>0.5$   \\  
2 & $3.02\pm 0.13$  & $0.83\pm 0.01$  & $0.14\pm 0.01$   & $-4.6\pm 0.5$  & $0.69\pm 0.01$   & $>0.7$   \\  
3 & $0.71\pm 0.07$  & $0.94\pm 0.01$  & $0.16\pm 0.01$   & $-3.9\pm 2.1$  & $0.75\pm 0.04$   & $1.2\pm 0.6$   \\  
4 & $0.33\pm 0.05$  & $1.22\pm 0.02$  & $0.26\pm 0.03$   & $0.6\pm 1.6$   & $0.72\pm 0.13$   & $1.0\pm 0.2$   \\ 
5 & $0.01\pm 0.03$  & $1.31\pm 0.05$  & $0.19\pm 0.11$   & $3.4\pm 4.8$   & $1.19\pm 0.30$   & $0.1\pm 0.6$   \\ 
6 & $0.14\pm 0.04$  & $1.37\pm 0.06$  & $0.29\pm 0.14$   & $-5.4\pm 3.2$  & $1.19\pm 0.28$   & $0.1\pm 0.6$   \\ 
7 & $0.04\pm 0.07$  & $1.27\pm 0.03$  & $0.00\pm 0.04$   & 0              & $1.23\pm 0.35$   & $0.1\pm 0.8$   \\ 
8 & $0.22\pm 0.05$  & $1.22\pm 0.02$  & $0.22\pm 0.04$   & 0              & $0.87\pm 0.27$   & $0.7\pm 0.7$   \\ 
\tableline \\[-1.0cm]
\end{tabular}
\tablecomments{All models are 3D. See \S \ref{dem} for details of the models. 
When a lower limit is given it represents the lowest value obtained
from 20 error simulations; i.e., it is essentially a 95\% confidence
lower limit.}
\end{center}
\end{table*}

Since the 2T models are preferred over the 1T models (particularly in
bins 1-4), we investigated other multitemperature spectral models to
attempt to constrain the general differential emission measure (DEM)
of the emission spectrum. Because the task of inferring the general
DEM from the X-ray spectrum of a coronal plasma is difficult and often
highly degenerate \citep{craig76}, especially for data of only
moderate energy resolution such as provided by the \xmm\ and \chandra\
CCDs, we probed the DEM using a few parameterized models. For this
discussion we focus on models fitted to the deprojected spectra; i.e.,
3D models.

\bigskip
\noindent{\it Cooling Flow:} The DEM for an ideal gas cooling at
constant pressure from a maximum temperature, \tmax, is
\citep[e.g.,][]{john92},
\begin{equation}
\xi_{\rm cf}(T) = {5\over 2}{\dot{M}k_B\over \mu m_p} {1 \over \Lambda(T)},
\hskip 0.5cm 
\label{eqn.cfdem} 
\end{equation}
for $T\le \tmax$. Here $k_B$ is Boltzmann's constant, $m_p$ is the
proton mass, $\Lambda(T)$ is the plasma emissivity which we take to be
given by the \apec\ code, and $\dot{M}$ is the mass drop-out rate.
The luminosity of this cooling gas emitted over some energy range
$\Delta E$ is,
\begin{equation}
L_{\Delta E}^{\rm cf} = \int^{\tmax}_{\tmin}\Lambda_{\Delta
E}(T)\xi_{\rm cf}(T)dT, \label{eqn.lum}
\end{equation}
where $\Lambda_{\Delta E}(T)$ is the plasma emissivity integrated over
the bandwidth ${\Delta E}$. We set $\tmin=0.05$~keV. The total
emission spectrum of a cooling flow model is taken to be $L_{\Delta
E}^{\rm cf}$ plus the spectrum of a $1T$ model with $T=\tmax$ to
represent the emission from the ambient gas. We shall refer to this
cooling flow model as ``CF+1T''. Note that this model adds only one
free parameter over the 1T case, i.e., $\dot{M}$, since we tie
together the abundances between the two components.

The $\chi^2$ values for the CF+1T model fitted to the \xmm\ and
\chandra\ data are listed in Table \ref{tab.fits.dem}. Although the
CF+1T fits are an improvement over the 1T models in some bins, the
improvement is not nearly as great as observed for the 2T model. The
largest improvement occurs in shells 3-4 where the $\chi^2$ values of
the CF+1T model are roughly half-way between the 1T and 2T values. The
CF+1T model offers no improvement over the 1T model in bins 1-2.

The total mass deposition rate (see Table \ref{tab.dem}) for all
shells (i.e., within $r=97$~kpc) is, $\mdot=4.5\pm 0.2$~\msunyr. This
value is less than half that obtained from the \rosat\ imaging
analysis of NGC 5044 by \citet{davi94} and only $\approx 1/8$ that
obtained from the single-aperture analysis of the \asca\ data by
\citet{buot99a}. The much smaller value of \mdot\ obtained from the
spatially resolved spectral analysis of the \xmm\ and \chandra\ data
is very similar to the reduction obtained from \chandra\ and \xmm\
results for cooling flow clusters \citep[e.g.,][]{davi01a}.

In fact, \mdot\ is even slightly smaller because the value in the
bounding shell is very likely overestimated. The spike in \mdot\ in
shell 8 is almost certainly biased by the edge effect inherent in our
deprojection procedure (\S \ref{deproj}).

\bigskip
\noindent{\it Gaussian:} An important consideration is whether the better
fits obtained by the 2T (3D) model over the 1T (3D) model could simply
reflect the range of temperatures within each shell arising from the
radially varying temperature profile implied by the 1T fits. In
support of this hypothesis is the fact that the improvement of the 2T
model is greatest in shells 1-4 where the 1T temperature profile is
changing most rapidly with radius; i.e., there we should have the
largest range of temperatures per shell.

To test this hypothesis we have explored a model where the temperature
distributions within each shell are a Gaussian. The Gaussian DEM
(GDEM) is expressed as,
\begin{equation}
\xi_{\rm g}(T) = \frac{norm}{\sigma_T\sqrt{2\pi}}\exp\left[ -{(T-T_0)^2 /
2\sigma_T^2} \right],
\end{equation}
where $T_0$ is the mean and $\sigma_T$ the standard deviation of the
Gaussian, and $norm$ is a constant proportional to $n_en_pV$ as
defined for the \apec\ code in \xspec\ (see caption to Figure
\ref{fig.em}). The luminosity over some energy range $\Delta E$ is
therefore given by equation (\ref{eqn.lum}) with $\xi_{\rm g}(T)$
replacing $\xi_{\rm cf}(T)$ and \tmax\ and \tmin\ set to $\pm
\infty$. 

In shells 1-4 the GDEM model provides superior fits to the 1T and
cooling flow models, but does not have $\chi^2$ values quite as low as
the 2T model. For example, in shell 3 we obtain for the Gaussian
model, $\chi^2=869.7$ for 692 dof compared to $\chi^2=1189.0$ for
1T, $\chi^2=982.1$ for CF+1T, and $\chi^2=836.0$ for 2T. Note that the
Gaussian adds only \sigmat\ as a free parameter over the 1T model;
i.e., it has the same number of free parameters as the cooling flow
model but one less than the 2T model.

The fitted values of \sigmat\ for shells 5-7 are consistent with the
single-phase hypothesis within the $1-1.5\sigma$ errors on
\sigmat. For shell 8 $\sigmat=0.22\pm 0.05$~keV could
reflect the projection of cooler gas from exterior shells as discussed
in \S \ref{2t} for the 2T models. We cannot determine this for certain
with the present data set.

In contrast, in shells 2-4 (and perhaps shell 1) the fitted values of
$T_0$ and \sigmat\ do not appear to be consistent with the radially
varying single-phase hypothesis. The $1\sigmat$ ranges for the
temperatures of the GDEM model are 0.68-0.97~keV for shell 2,
0.78-1.11~keV for shell 3, 0.95-1.48~keV for shell 4. The considerable
overlap of these $1\sigma$ values is in conflict with the small
temperature range expected from the single-phase hypothesis. That is,
for the 1T (3D) model, the temperature difference between shells 1 and
3 is $0.21\pm 0.01$~keV, between shells 2 and 4 is $0.32\pm 0.02$~keV,
and between shells 3 and 5 is $0.35\pm 0.02$~keV. The single-phase
hypothesis predicts, therefore, a range of temperatures of
approximately half the difference between adjacent shells; i.e.,
$\approx 0.11$~keV within shell 2, $\approx 0.16$~keV within shell 3,
and $\approx 0.17$~keV within shell 4. These values are considerably
smaller than the $1\sigmat$ ranges of $0.28\pm 0.02$~keV for shell 2,
$0.32\pm 0.02$~keV for shell 3, and $0.52\pm 0.06$~keV for shell 4.

The temperature difference between shells 1-2 is $0.09\pm 0.01$~keV
for the 1T (3D) model implying a single-phase temperature range of
$\approx 0.05$~keV. The $1\sigmat$ range for the GDEM model in shell 1
is $\sigmat=0.10\pm 0.04$~keV which is only marginally discrepant with
the single-phase hypothesis. 

\bigskip
\noindent{\it Power Law:} Finally, to further quantify deviations from the single-phase
hypothesis we have investigated spectral fits using a power-law DEM
(PLDEM),

\begin{equation}
\xi_{\rm pl}(T) = \begin{cases}
{norm \over \log (T_{\rm max}/ T_{\rm min})} T^{\alpha}, &
\alpha=-1\\[+5pt] {norm(\alpha + 1) \over (T_{\rm max}^{\alpha+1} -
T_{\rm min}^{\alpha+1})} T^{\alpha}, & \alpha\neq -1\\
\end{cases}
\label{eqn.pl}
\end{equation}
where $norm$ is a constant proportional to $n_en_pV$ as defined for
the \apec\ code in \xspec as discussed previously (see caption to
Figure \ref{fig.em}. The luminosity over some energy range $\Delta E$
is therefore given by equation (\ref{eqn.lum}) with $\xi_{\rm pl}(T)$
replacing $\xi_{\rm cf}(T)$. This model adds two free parameters over
the 1T model: $\alpha$ and the width of the temperature distribution,
$\tmax-\tmin$; i.e., the power-law model has the same number of free
parameters as the 2T model. Because the values of $\alpha$ and
$\tmax-\tmin$ are correlated we found it necessary to restrict their
values during the fits to $-5\le\alpha\le 5$ and $\tmax-\tmin \le
10$~keV.

As indicated in Table \ref{tab.fits.dem}, the PLDEM model fits nearly
as well as the 2T model in every shell. Because $\alpha$ is not well
constrained, particularly at larger radii, we fixed $\alpha\equiv 0$
in in shells 7-8. Even with this restriction, the power-law model fits
nearly as well as the 2T model in those shells.

The results obtained for $\alpha$, \tmin, and $\tmax-\tmin$ are listed
in Table \ref{tab.dem}. Like both the 2T and GDEM models, in shells
5-7 there is little indication of significant multitemperature gas;
i.e., $\tmin\ga 1.2$~keV and $\tmax-\tmin$ is small (though
uncertain). Similarly, in shell 8 the PLDEM model indicates the
presence of multiple temperature components which likely arises from
projection of gas exterior to shell 8 (see \S \ref{2t}).

For shells 1-4 the PLDEM fits imply a substantial range of temperature
components in each shell. In shell 4 we have $\alpha\approx 0$ which
indicates an equal contribution from temperature components over a
range from approximately 0.7~keV to 1.7~keV (with some uncertainty in
the upper limit). The value of $\alpha$ becomes increasingly negative
at smaller radius showing that the temperature distribution is
progressively more peaked near \tmin. However, the large values of
$\tmax-\tmin$ reflect the rise in \thot\ seen for the 2T models at
lower radius which is probably attributed to the emission from
unresolved stellar sources (\S \ref{2t}).

Overall, the PLDEM model, like the 2T and GDEM models, indicate that
within shells 1-4 the temperature distribution is too wide within each
shell to be described by a radially varying single-phase medium. The
\xmm\ CCD data cannot distinguish between the PLDEM and 2T models;
i.e., a continuous versus a discrete temperature distribution.

\subsection{Smaller Regions with Chandra}
\label{chan}

\begin{table*}[t] \footnotesize
\caption{$\chi^2$/dof for Selected Models Fitted Only to the \chandra\ Data
\label{tab.chan}} 
\begin{center} \vskip -0.4cm
\begin{tabular}{cccccc} \tableline\tableline\\[-7pt]
& $R_{\rm in}$ & $R_{\rm out}$\\ 
Shell & (arcmin) & (arcmin) & 1T & 2T & GDEM\\ 
\tableline \\[-7pt]
1C & 0.00 & 0.25  & 100.7/65  &   94.1/63   &  98.3/64 \\  
2C & 0.25 & 0.50  & 130.9/85  &   95.7/83   & 118.2/84 \\  
3C & 0.50 & 1.00  & 177.1/116 &  122.4/114  & 149.1/115\\  
4C & 1.00 & 1.50  & 196.5/120 &  129.4/118  & 141.5/119\\ 
5C & 1.50 & 2.00  & 235.1/118 &  159.1/116  & 164.7/117\\ 
6C & 2.00 & 2.50  & 170.1/116 &  138.9/114  & 148.0/115\\
\tableline \\[-1.0cm]
\end{tabular}
\tablecomments{All models are 3D. These models also account for
emission projected from shells outside shell 6C using the results
obtained from fitting the \xmm\ data outside shell 6C.}
\end{center}
\end{table*}

\begin{table*}[t] \footnotesize
\caption{Parameters for Selected Models Fitted Only to the \chandra\ Data
\label{tab.chan.pars}} 
\begin{center} \vskip -0.4cm
\begin{tabular}{cc|cc|cc} \tableline\tableline\\[-7pt]
& 1T & \multicolumn{2}{c}{2T} & \multicolumn{2}{c}{GDEM}\\
& $T$ & \tcool & \thot & $T_0$ & \sigmat \\
Shell & (keV) & (keV) & (keV) & (keV) & (keV)\\ 
\tableline \\[-7pt]
1C & $0.61\pm 0.04$ & $0.61\pm 0.03$  & $\cdots$       & $0.63\pm 0.11$  &$0.15\pm 0.14$	      \\  
2C & $0.73\pm 0.03$ & $0.70\pm 0.03$  & $2.1\pm 0.6$    & $0.77\pm 0.05$  &$0.23\pm 0.08$	      \\  
3C & $0.75\pm 0.01$ & $0.75\pm 0.01$  & $2.2\pm 1.2$   & $0.76\pm 0.01$  &$0.11\pm 0.05$	      \\[+3pt]  
4C & $0.79\pm 0.01$ & $0.78\pm 0.02$  & $1.58\pm 0.35$ & $0.84\pm 0.01$  &$0.14_{-0.02}^{+0.29}$\\[+3pt]  
5C & $0.88\pm 0.01$ & $0.80\pm 0.03$  & $1.45\pm 0.13$ & $0.91\pm 0.03$  &$0.19_{-0.09}^{+0.24}$\\[+3pt]  
6C & $0.91\pm 0.01$ & $0.84\pm 0.02$  & $1.27\pm 0.23$ & $0.93\pm 0.03$  &$0.10_{-0.03}^{+0.30}$\\
\tableline \\[-1.0cm]
\end{tabular}
\tablecomments{Models correspond to those in Table
\ref{tab.chan}. Quoted errors using the symbol ``$\pm$'' are
$1\sigma$. Since the error ranges of \sigmat\ for shells 4C-6C are very
asymmetric we quote the extreme values obtained from the 20 Monte
Carlo error runs; i.e., these are effectively 95\% confidence limits.}
\end{center}
\end{table*}

\begin{figure*}[t]
\parbox{0.49\textwidth}{
\centerline{\psfig{figure=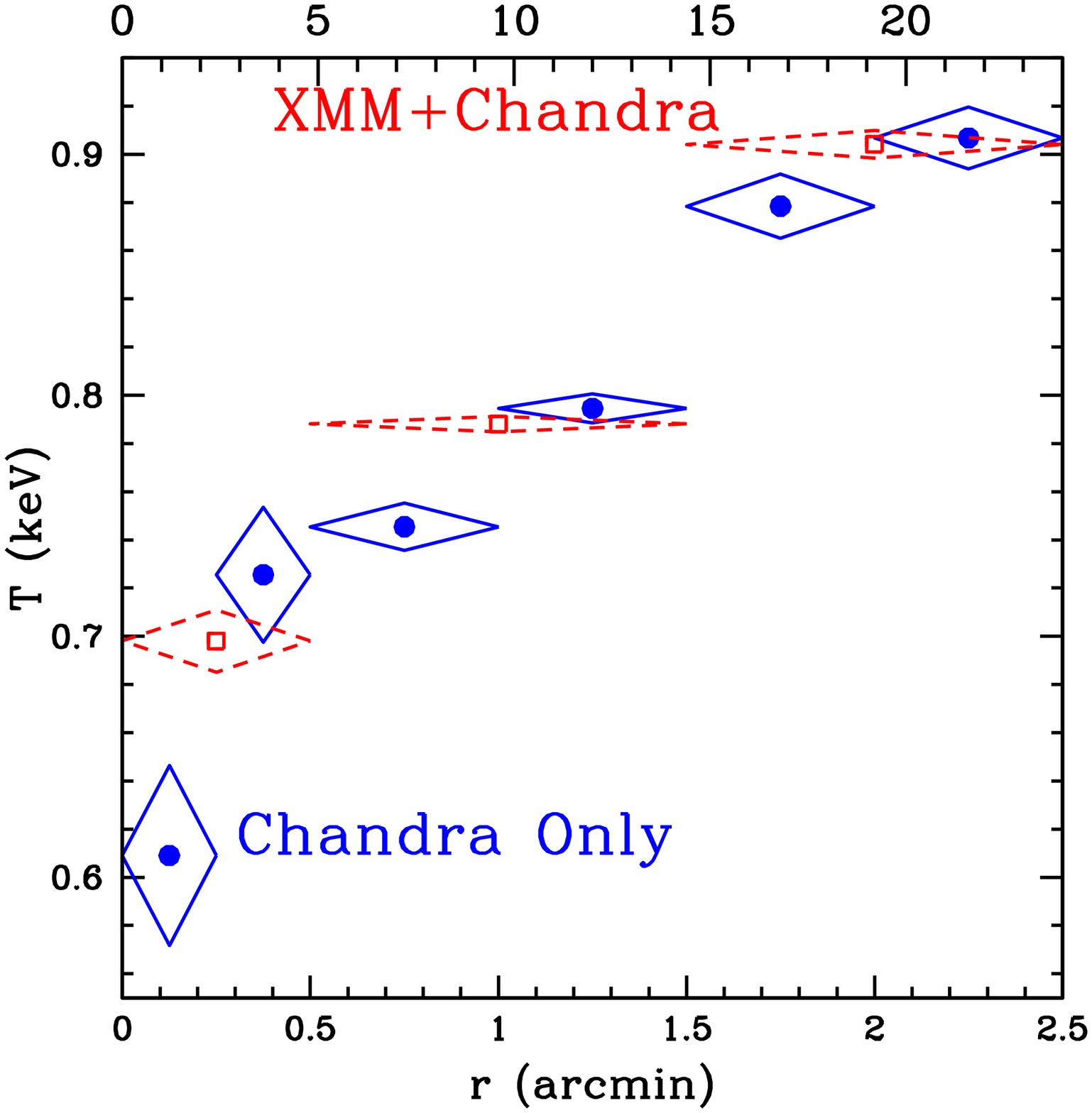,height=0.35\textheight}}
}
\parbox{0.49\textwidth}{
\centerline{\psfig{figure=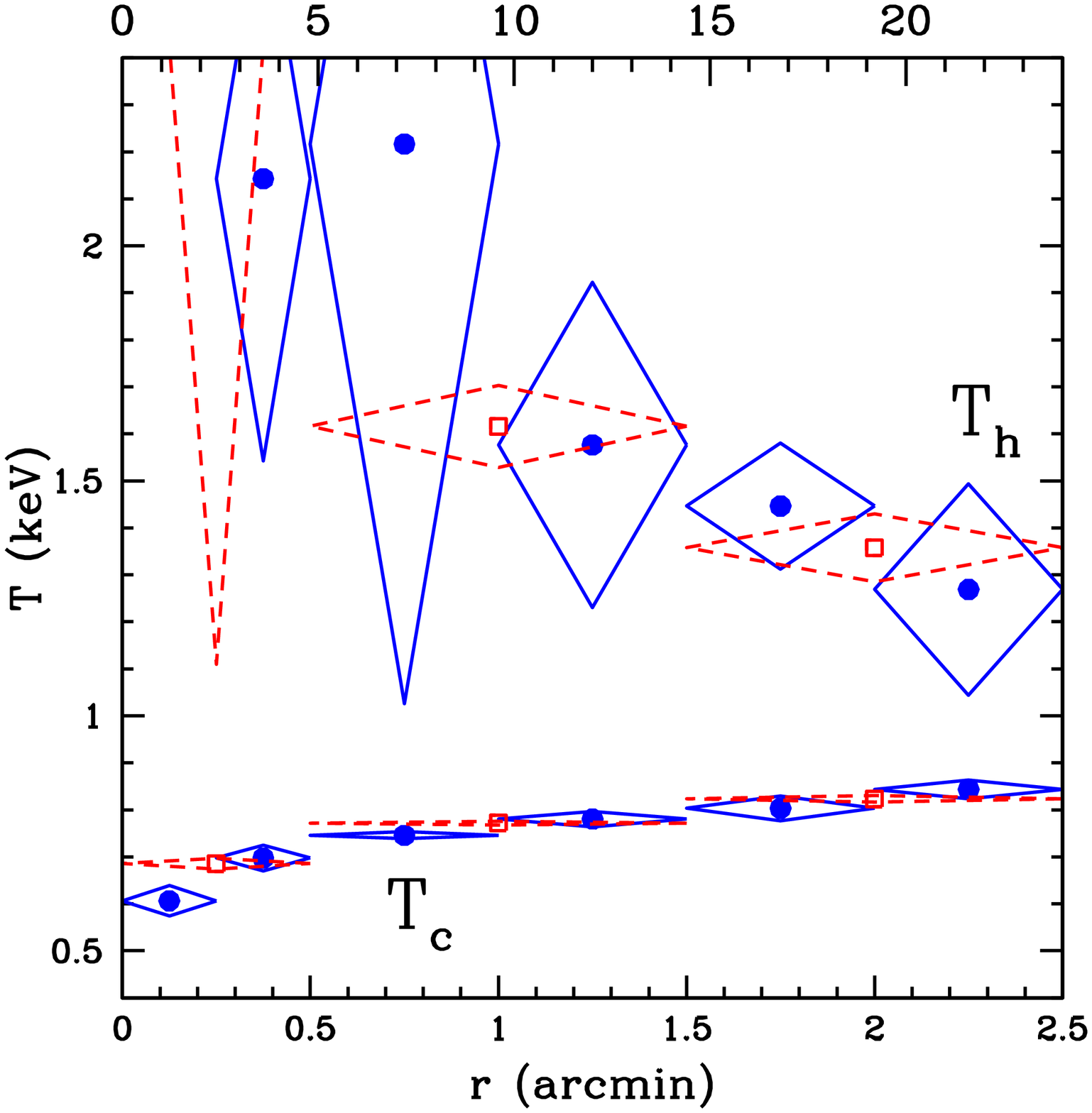,height=0.35\textheight}}
}
\caption{\label{fig.temp.chan} 3D radial temperature profiles
(units -- bottom: arcminutes, top: kpc) and $1\sigma$ errors for ({\sl
Left panel}) 1T and ({\sl Right panel}) 2T models. The (blue) circles
and solid diamonds refer to models fitted only to the \chandra\ data
in shells 1C-6C.  The (red) boxes and dashed diamonds refer to models
fitted jointly to the \xmm\ and \chandra\ data in shells 1-3.}
\end{figure*} 

The higher spatial resolution of \chandra\ allows for smaller regions
to be probed than with \xmm. This allows the preference for
multitemperature models over the radially varying single-phase
hypothesis of NGC 5044 to be tested with \chandra\ using shells of
smaller width. If the hot gas is really a radially varying
single-phase medium, then the temperature range indicated by a 2T
model should decrease with decreasing aperture size.  Therefore, we
have further divided the region spanned by shells 1-3 into the six
shells 1C-6C defined in Table \ref{tab.chan}. We are only able to
usefully probe apertures of about half the size used previously
because the \chandra\ ACIS-S3 data alone do not have the combined
sensitivity of the \xmm\ CCDs.

In Table \ref{tab.chan} we list the $\chi^2$ values for the 1T and 2T
(both 3D) fits in shells 1C-6C. In good agreement with the results
presented above, we find that the 2T model is a better fit than the 1T
model. The temperatures of the 1T and 2T models are plotted in Figure
\ref{fig.temp.chan} and listed in Table \ref{tab.chan.pars}. The 1T
profile shows excellent agreement with the results obtained from the
joint \chandra-\xmm\ fits. In particular, notice that it is the
temperature values obtained for shells 2C, 4C, and 6C that agree well
with those obtained from the larger shells. The lower temperature
values obtained in 1C, 3C, and 5C indicate that the emission-weighted
temperatures obtained in shells 1-3 are heavily weighted toward the
outer edge of the shell.

The temperatures for the 2T model in the smaller \chandra\ apertures
also agree well with those obtained from the larger apertures. The
value of \thot\ in shell 1C is not shown since it has a large
best-fitting value and is effectively unconstrained. Note also that
results for a small number of error simulations where the
normalization of \thot\ in shells 2C and 3C is very small are not
included in the error bars indicated in the figure.

It is clear that the 2T models fitted to the \chandra\ data alone
indicate a range of temperatures at a given radius consistent with the
2T models fitted within wider shells. In shells 4C-6C the temperature
is consistent with a constant value, $\thot$ between 1.3-1.5~keV. The
temperature difference within each shell is significantly larger than
indicated by the radially varying single-phase model. For example, in
shell 5C where the ratio the emission measures for the cool and hot
components is 2.3 (best fit), we have $\thot-\tcool=0.65\pm 0.13$~keV
which is substantially larger ($5\sigma$) than the $\approx
0.03$~keV spread in that shell expected from the single-phase
hypothesis.

Perhaps a clearer, more quantitative, indication of the temperature
range within the shells is indicated by the GDEM model.  The $\chi^2$
values and temperature parameters are listed in Tables \ref{tab.chan}
and \ref{tab.chan.pars} are for the GDEM model. Similar to the results
obtained from the joint \xmm--\chandra fits in larger regions (\S
\ref{dem}), the $\chi^2$ values obtained for the GDEM model are
generally intermediate between the 1T and 2T models, though are
clearly better than 1T in all shells except 1C. The parameter values
for the GDEM model obtained from the \chandra\ data alone are
consistent with those obtained in the wider apertures (cf. Table
\ref{tab.dem}) though with larger error bars.

As we did in \S \ref{dem} we can use the GDEM model to test the
hypothesis that the temperature profile of the hot gas represents a
radially varying single-phase medium. Using the 1T results for shells
3C-6C we would expect the following range of temperatures within
shells 4C-6C: $\approx 0.03$~keV within 4C and 5C, and $\approx
0.02$~keV within 6C. Since the 95\% confidence lower limit on \sigmat\
is $\approx 0.10$ in these bins (Table \ref{tab.chan.pars}), the
$1\sigmat$ temperature range implied by the GDEM model within each
shell 4C-6C is at least 0.20~keV -- very much larger than than the
0.02-0.03~keV ranges expected from the single-phase
hypothesis. (Consistent evidence for multiphase gas is indicated in
shells 2C and 3C, but the significance is only at the 2-3 $\sigma$
level.)

We conclude that the improvement in the fits provided by the GDEM and
2T models, as well as the large implied temperature widths obtained
for the \chandra\ data in the thinner shells, provides important
additional evidence that a single-phase description of the hot gas in
NGC 5044 is inadequate in the central regions.

\section{Non-Radial Analysis}
\label{2d}

\subsection{XMM}

We performed a two-dimensional spectral analysis to determine whether
the multiple components inferred from the analysis assuming spherical
symmetry arise from azimuthal temperature fluctuations within each
annulus.  We searched for azimuthal variations in the temperature and
metal abundances using the following simple procedure; results
obtained for the abundances are discussed in Paper 2. Within the
central CCD of the MOS images we defined a 5x5 array of equally spaced
circular extraction regions of $1\arcmin$ radius. Just outside of the
central CCD, we defined 12 equally spaced circular regions of
$2\arcmin$ radius that surrounded the central CCD. Similar to the
azimuthally averaged analysis, to each region we fitted 1T and 2T
models modified by foreground Galactic absorption. Each model was
fitted simultaneously to the MOS1 and MOS2 data projected on the sky;
i.e., no deprojection.

Overall, we find results consistent with the spherically symmetric
analysis. The temperatures obtained from the 1T model generally vary
significantly only with distance from the center of the
image. However, we notice a small asymmetry in the temperature
distribution at a radius between $2\arcmin-3\arcmin$. At this radius
we find $T\approx 0.9$~keV for $\theta=90\degr-180\degr$ while
$T\approx 1.1$~keV elsewhere at this radius ($\theta$ is measured
N-E.). This small azimuthal variation cannot account for the wide
temperature distributions inferred from the 2T and continuous DEM
models. 

In fact, when 2T models are fitted to the data, results in very good
agreement with those obtained from the spherical analysis are
obtained. The fits are improved over the 1T models and values for
\tcool\ and \thot\ are obtained consistent with those presented for
the 2T (2D) models in \S \ref{2t}.

\subsection{\chandra}
\label{2d_chan}

\begin{table*}[t] \footnotesize
\caption{Parameters from the Non-Radial Analysis of the \chandra\ Data
\label{tab.chan_2d.pars}} 
\begin{center} \vskip -0.4cm
\begin{tabular}{ccccccc}
\tableline\tableline\\[-7pt]
& \multicolumn{2}{c}{1T} & \multicolumn{4}{c}{2T}\\
& & $T$ & & \tcool & \thot & n$_c$/n$_h$ \\
Region & $\chi^2$/dof & (keV) & $\chi^2$/dof & (keV) & (keV) & (ratio) \\
\tableline \\[-7pt]
{\emph{Annulus 2}} \\
1A &   115.6/79  & $0.82 \pm 0.005$ & 81.7/77  & $ 0.80 \pm 0.009 $ & $1.84 \pm 0.22 $ & $3.0 \pm 0.6$ \\  
1B &   58.3/46   & $0.76 \pm 0.009$ & 53.8/44  & $ 0.76 \pm 0.022 $ & \nodata 	     & \nodata 	     \\  
2  &   114.3/83  & $0.78 \pm 0.005$ & 71.8/81  & $ 0.74 \pm 0.015 $ & $1.42 \pm 0.13 $ & $3.0 \pm 0.7$ \\  
3A &   118.0/82  & $0.81 \pm 0.005$ & 79.6/80  & $ 0.78 \pm 0.010 $ & $1.52 \pm 0.19 $ & $3.3 \pm 0.8$ \\  
3B &   42.5/42   & $0.82 \pm 0.010$ & 35.0/40  & $ 0.82 \pm 0.015 $ & \nodata 	     & \nodata       \\  
{\emph{Annulus 3}} \\
4  &   47.6/41   & $1.07 \pm 0.018$ & 45.9/39  & $ 1.06 \pm 0.059 $ & \nodata 	     & \nodata       \\  
5  &   65.3/40   & $1.12 \pm 0.031$ & 55.8/38  & $ 1.04 \pm 0.067 $ & \nodata  	     & \nodata       \\  
6A &   80.4/61   & $0.77 \pm 0.013$ & 54.8/59  & $ 0.73 \pm 0.017 $ & $1.56 \pm 0.21 $ & $2.9 \pm 0.8$ \\  
6B &   33.1/31   & $0.75 \pm 0.017$ & 26.6/29  & $ 0.67 \pm 0.038 $ & $1.14 \pm 0.19 $ & $2.1 \pm 1.1$ \\  
\tableline \\[-1.0cm]
\end{tabular}
\tablecomments{Models correspond to those in Table
\ref{tab.chan}.  Regions correspond to Figure \ref{fig.2d.chan}.
Regions with no entry for \thot\ did not significantly constrain that
parameter.  Quoted errors using the symbol ``$\pm$'' are
$1\sigma$. The final column is the ratio of emission measures of the
cold and hot temperature components.}
\end{center}
\end{table*}

\begin{figure*}[t]
\centerline{\psfig{figure=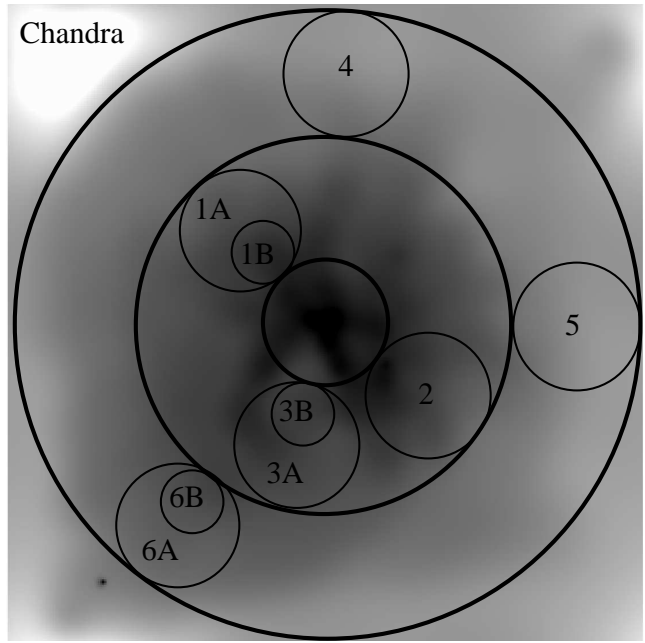,height=0.35\textheight}}
\caption{\label{fig.2d.chan} Portion of the archival \chandra\ ACIS-S image
located at the center of NGC 5044 adaptively smoothed using the CIAO task
{\sc CSMOOTH}. Heavy circles mark the limits of annuli 2 and 3. 
Overlaid are several regions used in our non-radial analysis (see
\S \ref{2d_chan}). Region labels match Table \ref{tab.chan_2d.pars}.}
\end{figure*}

We have performed a similar but independent two-dimensional analysis on the
\chandra\ data. The higher spatial resolution of \chandra\ allows a more
definitive test of the existence of large-scale multiple temperature
gas components, by allowing regions as small as a few arcseconds in
radius to be analyzed in principle. To directly compare with the
azimuthally-averaged analysis, we have chosen circular regions which
fall within annuli 2 and 3 defined in Table \ref{tab.fits}. In
Fig. \ref{fig.2d.chan} we present several representative regions from
this analysis overlaid on a smoothed \chandra\ image (we note that we
have analyzed similar regions across the entire image, but do not find
further variation beyond what we report below). Six of these
representative regions are 30\arcsec\ in radius, thus reaching the
edge of the annulus they fall within. In addition, we have chosen
three smaller regions 15\arcsec\ in radius which lie within the larger
regions, and would fall within the thinner annuli 3C and 5C (defined
in Table \ref{tab.chan}). Some of the regions were placed
intentionally in the areas of bright or faint emission visible in the
smoothed image, which might be expected to exhibit the strongest
spectral variations. As above, in each region we fitted 1T and 2T
models modified by foreground Galactic absorption.

In Table \ref{tab.chan_2d.pars} we present the results of the 1T and
2T fits for the selected regions shown in Fig. \ref{fig.2d.chan}. We
again find results entirely consistent with the spherically symmetric
analysis. For the 1T model, the temperatures from regions lying within
annulus 2 are clearly grouped about the value obtained for the entire
annulus (cf. Figure \ref{fig.temp}), not deviating enough to explain
the multi-temperature results of the annular analysis. In annulus 3,
we see the same asymmetry as observed in the \xmm\ non-radial
analysis, with $T\approx 0.8$~keV for $\theta=90\degr-180\degr$ while
$T\approx 1.1$~keV elsewhere at this radius. Again, this variation
cannot account for the wide temperature distributions inferred from
the 2T and continuous DEM models.

The 2T model fits provide an impressive confirmation of this
result. For regions 1A, 2, 3A, 6A, and 6B, the 2T models are
significantly improved over the 1T models, and obtain values for
\tcool\ and \thot\ consistent with those fit to the entire annulus (\S
\ref{2t}). The strongest constraints are within annulus 3, which is
expected since the emission measures of the hotter and cooler
components are most similar there (\S \ref{2t}; see Figure
\ref{fig.em}). For the four remaining regions, there is insufficient
S/N to constrain the value of a second temperature component. In these
regions (1B, 3B, 4, 5), a second temperature component of similar
temperature and relative emission measure strength as found in other
regions can be added without significant effect on the fit. We
emphasize that regions 1B and 3B are simply subsets of their
surrounding regions 1A and 3A, wherein a 2T model improved the
fit. Presumably, in a longer exposure of the smaller regions, the same
2T result would obtain; i.e., while present, the 2T model is simply
unconstrained in the low S/N regime.

Spatial variations in temperature could also be investigated with a
hardness ratio map, an analysis which has been performed by
\citet{tamu03a}. They do not find any significant azimuthal hardness
variations, consistent with our non-radial analysis here.

\section{Systematic Errors}
\label{sys}

This section contains a detailed investigation of systematic errors on
the temperature measurements. Those readers who are not interested in
these technical details can safely skip ahead to \S \ref{bigap}.

\subsection{Calibration}

\subsubsection{XMM--EPIC vs. Chandra--ACIS}

\begin{table*}[t] \footnotesize
\caption{Comparison of Temperatures from XMM and Chandra
\label{tab.xmmchan}} 
\begin{center} \vskip -0.4cm
\begin{tabular}{cccc|cccccc} \tableline\tableline\\[-7pt]
& \multicolumn{3}{c}{1T} & \multicolumn{6}{c}{2T}\\ \tableline \\[-7pt]
& \multicolumn{3}{c}{$T$} & \multicolumn{3}{c}{\tcool} & \multicolumn{3}{c}{\thot}\\
Annulus & XMM & Chandra & \% & XMM & Chandra & \% & XMM & Chandra & \% \\  
\tableline \\[-7pt]
1 & $0.764\pm 0.005$ & $0.738\pm 0.009$ & $3\pm 1$ & $0.747\pm 0.006$ & $0.724\pm 0.010$ & $3\pm 2$ & $1.62\pm 0.11$ & $2.08\pm 0.31$ & $-28\pm 20$\\
2 & $0.814\pm 0.004$ & $0.794\pm 0.004$ & $3\pm 1$ & $0.790\pm 0.002$ & $0.782\pm 0.005$ & $1\pm 1$ & $1.46\pm 0.04$ & $1.54\pm 0.07$ & $-5\pm 6$\\
3 & $0.990\pm 0.004$ & $0.975\pm 0.005$ & $2\pm 1$ & $0.830\pm 0.005$ & $0.835\pm 0.008$ & $-1\pm 1$ & $1.36\pm 0.03$ & $1.41\pm 0.06$ & $-4\pm 5$\\
\tableline \\[-1.0cm]
\end{tabular}
\tablecomments{All models are 2D to allow an independent comparison of
the data sets in the central regions. The temperature is expressed in
keV. ``\%'' is the percent difference between the \xmm\ and
\chandra\ temperatures.}
\end{center}
\end{table*}

We have examined possible systematic errors in the measurements of the
temperatures arising from calibration differences between the \xmm\
and \chandra\ CCDs. In Table \ref{tab.xmmchan} we list the
temperatures obtained from 1T (2D) and 2T (2D) models fitted
separately to the \xmm\ and \chandra\ data; i.e., the MOS and pn data
were fitted simultaneously while the \chandra\ data were fitted
alone. We focus on 2D models so that the fits for a specific annulus
are independent of results obtained from fits to adjacent regions at
larger radii.

Using the 1T model the \xmm\ and \chandra\ data in annuli 1 and 3 give
values of \fe\ that agree within 3\% and are consistent within their
$3\sigma$ statistical errors. The temperatures of the 2T model are
more uncertain, and the \xmm\ and \chandra\ temperatures are
consistent within their $1-1.5\sigma$ errors.

\subsubsection{EPIC and ACIS CCDs vs. RGS Gratings}

We have compared results obtained from our analysis of NGC 5044 using
the \xmm\ and \chandra\ CCDs to the published results obtained for the
\xmm\ RGS obtained by \citet{tamu03a}. First, by comparing the
emission measures obtained within our extraction radii with the
published emission measure for the RGS data by  \citet{tamu03a} and a
separate analysis of the RGS data (T. Fang 2003, private
communication) we conclude that the effective RGS aperture corresponds
to a circular aperture with radius $R\approx 30\arcsec - 40\arcsec$;
i.e., essentially our radius bin 1. 

\citet{tamu03a} find that a 2T (2D) model is a significant improvement over
a 1T model. In their analysis they (1) restrict their analysis to
0.44-1.55~keV, (2) only allow the iron and oxygen abundances to vary,
(3) use the solar abundance table of \citet{angr}, and (4) use a
plasma code essentially corresponding to the \mekal\ code
\citep{meka,mekal} in \xspec. They obtain the following parameters for
the 2T model: $\tcool=0.7$~keV (no error quoted), $\thot=1.07\pm
0.03$~keV, $\fe=0.55\pm 0.05\solar$, and $\ox/\fe = 0.51\pm 0.06$ in
solar units.  If we follow the same procedures (1)-(4) for the EPIC
and ACIS data within radial bin 1 we obtain the following best-fitting
values for the 2T model: $\tcool=0.68$, $\thot=1.08$~keV,
$\fe=0.58\solar$, and $\ox/\fe=0.46$ in solar units. These results are
in excellent agreement with Tamura et al.'s results for the RGS. 

We conclude that the RGS and the EPIC and ACIS CCDs give fully
consistent results when the same models are fitted over the same
energy ranges for each instrument. In particular, the RGS corroborates
the improvement of a 2T (2D) model over a 1T (2D) model. However,
since they are unable to perform a deprojection analysis,
\citet{tamu03a} could not determine whether the extra temperature
component arises only from projection of gas from larger radii. (The
consistency of the metal abundances is discussed further in Paper 2.)

\subsection{Plasma Codes} 
\label{plasma}

We compared the results obtained using the \apec\ code to those
obtained using the \mekal\ code \citep{meka,mekal} to assess the
importance of different implementations of the atomic physics and
different emission line lists in the plasma codes. In every case
examined we found no qualitative differences between results obtained
from each code; e.g., the fitted temperatures usually agree to within
$\sim 5\%$. The $\chi^2$ values obtained for the multitemperature
models are also very similar. Some significant quantitative
differences in $\chi^2$ values are observed for the 1T models, but
there is no qualitative difference in the fits. For example, for shell
3 the 1T (3D) \mekal\ model gives $\chi^2=1535.3$ for 693 dof compared
to $\chi^2=1338.2$ for the \apec\ code. Visual inspection of these
fits and residuals reveals no noticeable differences between the two
fits except for slightly more pronounced residuals in the Fe L region
in the \mekal\ fit.

(We note that we explored the the validity of the assumption of
ionization equilibrium using the {\sc vnei} model in \xspec. We found
no improvement in the fit for a single-temperature model when allowing
for departures from ionization equilibrium.)

\subsection{Bandwidth} 

We explored the sensitivity of our results our default lower limit of
the bandpass, $\emin=0.5$~keV. For comparison we performed 1T and 2T
fits (both in 3D) with $\emin=0.3$~keV and $\emin=0.7$~keV. The fitted
temperatures are generally consistent between models with \emin\
between 0.3-0.7~keV. However, the $\chi^2$ values indicate that the
improvement of the 2T model over the 1T model decreases significantly
as \emin\ increases. In shell 3, for example, we obtain $\chi^2$/dof
of 1427.9/774 (1T) and 957.5/772 (2T) for $\emin=0.3$~keV , 1189.0/693
(1T) and 836.0/691 (2T) for $\emin=0.5$~keV, 916.4/614 (1T) and
730.5/612 (2T) for $\emin=0.7$~keV (each model is 3D).  Similar
behavior is observed for shells 1,3, and 4.

This decreasing of the need for multitemperature models with
increasing \emin\ demonstrates that the residuals in the Fe L
lines near 1~keV for 1T models are not the sole driving force for the
2T and other multitemperature models. {\bf This is significant since
it implies that remaining inaccuracies in the the Fe L lines in the
plasma codes cannot be solely responsible for the improvement of the
multitemperature models over 1T models.} Conversely, large bandwidth
is seen to be imperative in the search for, and constraint of,
multitemperature models of the hot gas.

\subsection{Variable \nh}
\label{nh} 

\begin{table*}[t] \footnotesize
\caption{1T and 2T Models with Variable Absorption
\label{tab.nh}} 
\begin{center} \vskip -0.4cm
\begin{tabular}{c|cc|cc} \tableline\tableline\\[-7pt]
& \multicolumn{2}{c}{1T} & \multicolumn{2}{c}{2T}\\
& & $\Delta\nh$ & & $\Delta\nh$\\
Shell & $\chi^2$/dof & ($10^{20}$~\cmsq) & $\chi^2$/dof & ($10^{20}$~\cmsq) \\ \tableline \\[-7pt]
1 &   688.7/450  & $5.2\pm 0.8$     & 637.8/448  & $1.9\pm 0.9$  \\  
2 &  1059.8/688  & $7.0\pm 0.4$     & 916.7/686  & $3.4\pm 0.6$  \\  
3 &  1120.9/692  & $4.2\pm 0.4$     & 828.2/690  & $1.9\pm 0.6$  \\  
4 &   616.3/481  & $4.0\pm 0.7$     & 533.6/479  & $1.3\pm 0.8$  \\ 
5 &   510.3/453  & $0.8\pm 0.7$     & 485.2/451  & $0.4\pm 0.8$  \\ 
6 &   551.4/468  & $0.8\pm 0.8$     & 513.8/466  & $0.4\pm 1.0$  \\ 
7 &   561.2/514  & $-1.8\pm 0.8$    & 544.7/512  & $-2.4\pm 0.9$ \\ 
8 &   674.3/560  & $0.5\pm 0.7$     & 657.9/558  & $0.8\pm 1.0$  \\ 
\tableline \\[-1.0cm]
\end{tabular}
\tablecomments{$\Delta\nh$ is the difference between the fitted
absorption column density and the assumed Galactic value ($5\times
10^{20}$~\cmsq). Both emission models are 3D while the absorption
model is a conventional foreground screen. See \S \ref{nh} for
description of the models.}
\end{center}
\end{table*}

Since the lever-arm provided by the bandwidth below $\sim 0.7$~keV is
an important constraint on the multitemperature models, it should be
expected that the fits may also be improved -- to some extent -- by
allowing for intrinsic (continuous) photoelectric absorption from cold
gas. In Table \ref{tab.nh} we present results for the 1T (3D) and 2T
(3D) models where we have allowed \nh\ of the foreground absorber
component to be a free parameter. (We have examined a suite of cold
absorber models and they give results similar to those for the simple
foreground-screen cold absorber model.) In shells 1-2 the 1T model is
improved significantly with variable absorption. However, in shells
1-4 the 2T model is still clearly preferred over the 1T model when
allowing for variable absorption in each case. In fact, variable \nh\
actually improves the 2T model very little; i.e., the 2T model with
Galactic absorption (Table \ref{tab.fits}) is itself superior to the
1T model with variable \nh in shells 1-4.

The superior fits provided by the 2T (and other multitemperature
models like the PLDEM) with Galactic absorption, and the relatively
insignificant improvement to the multitemperature models obtained
when intrinsic absorption is allowed for, indicates to us that there
is little motivation to consider intrinsic absorption
models. Moreover, the variable-\nh models imply large amounts of
absorbing material in shells 1-4; e.g., taking $\Delta\nh=2.5\times
10^{20}$~\cmsq\ in shells 1-2 leads to an absorbing mass, $M_{\rm
abs}\approx m_{\rm H}\Delta\nh\pi (14\, \rm kpc)^2 \approx
10^{9}\msun$ assuming solar abundances and that the absorber is
uniformly distributed; this is a lower limit if the abundances are
sub-solar and the absorber is non-uniform. This large absorbing mass
is almost as large as that of the hot gas. Yet this amount of cold
material, whether it be from optical line emitting gas or dust
\citep[e.g.,][]{goud94} or from atomic or molecular gas
\citep[e.g.,][]{breg92a,odea94a}, has never been seen in NGC 5044 or
other ``cooling flow'' galaxies. (We note that the fitted values of
\nh\ for both the 1T and 2T models at large radius are consistent with
the Galactic value within their $1-2\sigma$ errors.)

Since the multitemperature models with $\nh=\nhgal$ provide better
fits within the central $\approx 30$~kpc, there is no obvious sharp
absorption feature in the spectrum, and there is no evidence from
observations in other wavebands for the large quantities of cold
absorbing material implied by the fitted values of \nh, we do not take
seriously the results obtained from the intrinsic cold absorber
models.  

\subsection{Background} 

Since NGC 5044 is sufficiently bright the fitted temperature values
are quite insensitive to errors in the background normalization. For
example, even if we do not subtract the background in the bounding
annulus (where background is most important) the temperature we obtain
for a 1T model differs only by 10\% from the background-subtracted
value. Within the inner shells the background effect is negligible;
e.g., in shell 2 if the background is not subtracted we obtain a 1T
temperature that differs by $0.2\%$ from the background-subtracted
value.

\section{Spectral Analysis within a Large Aperture}
\label{bigap}

\begin{figure*}[t]
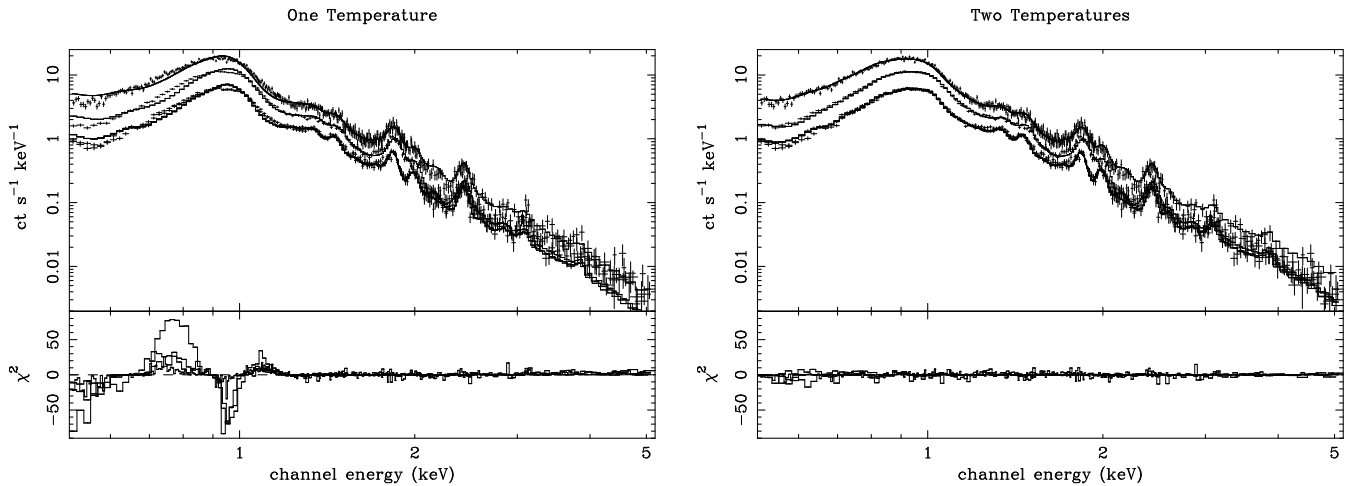

\parbox{0.49\textwidth}{
\centerline{\psfig{figure=f9a.eps,angle=-90,height=0.26\textheight}}}
\parbox{0.49\textwidth}{
\centerline{\psfig{figure=f9b.eps,angle=-90,height=0.26\textheight}
}}
\caption{\label{fig.bigap} EPIC MOS1, MOS2, pn and ACIS-S3 spectra
accumulated within a circular aperture of radius, $R=2.5\arcmin$
(24~kpc), fitted with ({\sl Left panel}) a single temperature (1T)
model and ({\sl Right panel}) a two-temperature (2T) model (no
deprojection is performed). In each case the \apec\ plasma model is
used and the solar abundances are taken from \citet{grsa} which use
the new (smaller) photospheric value for the iron abundance. The
models are the same as discussed in \S \ref{1t} and \S \ref{2t} for
the spatially resolved analysis. That is, Galactic absorption is
assumed and the following metal abundances are free parameters: O, Ne,
Mg, Si, S, Fe and all other abundances are tied to Fe in their solar
ratios. For the 2T model the abundances of each temperature component
are tied together in the fits.}
\end{figure*} 

\begin{table*}[t] \footnotesize
\caption{Selected Results for Models Fitted Within a Large Aperture
\label{tab.bigap}} 
\begin{center} \vskip -0.4cm
\begin{tabular}{lccccccc} \tableline\tableline\\[-7pt]
& & $\Delta\nh$ & \tcool & \thot &  & $T_{\rm min}$ & $T_{\rm max} - T_{\rm min}$\\
\multicolumn{1}{c}{Model} & $\chi^2$/dof & ($10^{20}$~\cmsq) & (keV) & (keV) & $\alpha$ & (keV) & (keV)\\
\tableline \\[-7pt]
1T, Galactic \nh & 4197.4/869 & $\cdots$ & $0.897\pm 0.002$ & $\cdots$ & $\cdots$ & $\cdots$ & $\cdots$\\
1T, variable \nh & 3879.5/868 & $4.5\pm 0.3$ & $0.878\pm 0.002$ & $\cdots$ & $\cdots$ & $\cdots$ & $\cdots$\\
2T, Galactic \nh & 1356.8/867 & $\cdots$ & $0.792\pm 0.002$ & $1.41\pm 0.02$ & $\cdots$ & $\cdots$ & $\cdots$\\
2T, variable \nh & 1313.0/866 & $3.0\pm 0.6$ & $0.779\pm 0.006$ & $1.28\pm 0.04$ & $\cdots$ & $\cdots$ & $\cdots$\\
PLDEM, Galactic \nh & 1373.0/867 & $\cdots$ & $\cdots$ & $\cdots$ & $-3.0\pm 0.1$ & $0.681\pm 0.005$ & $1.5\pm 0.1$\\
\tableline \\[-1.0cm]
\end{tabular}
\tablecomments{The (2D) models are fitted to the accumulated \xmm\ EPIC
MOS1, MOS2, pn and \chandra\ ACIS-S3 spectral data within a circular
aperture of radius $2.5\arcmin$ (24~kpc). $\Delta\nh$ is the
difference between the fitted absorption column density and the
assumed Galactic value. The emission measures for the 2T models are,
in the \xspec\ units discussed in the caption to Figure \ref{fig.em}:
$\rm norm_{\rm c} = 6.7$e-3 $\pm$ 1e-4 and $\rm norm_{\rm h} =
3.7$e-3~$\pm$~1e-4 for 2T, Galactic \nh\ and $\rm norm_{\rm c} =
8.1$e-3~$\pm$~3e-4 and $\rm norm_{\rm h} = 4.6$e-3~$\pm$~4e-4 for 2T,
variable
\nh.}
\end{center}
\end{table*}

\begin{table*}[t] \footnotesize
\caption{Fits to Simulated Data of Radially Varying Models Within a Large Aperture
\label{tab.bigap.sims}} 
\begin{center} \vskip -0.4cm
\begin{tabular}{l|ccc} \tableline\tableline\\[-7pt]
& \multicolumn{3}{c}{$\chi^2$/dof}\\
\multicolumn{1}{c}{Model} & Sim \#1 &  Sim \#2 &  Sim \#3\\ \tableline \\[-7pt]
1T, Galactic \nh &  1809.8/839 & 2469.2/855 & 3731.6/861\\
1T, variable \nh &  1805.3/838 & 1309.1/854 & 3611.3/860\\
2T, Galactic \nh &  865.6/837  & 1105.4/853 & 980.0/859\\
\tableline \\[-1.0cm]
\end{tabular}
\tablecomments{The models are fitted to simulated \xmm\ EPIC
MOS1, MOS2, pn and \chandra\ ACIS-S3 spectral data accumulated within
a circular aperture of radius $2.5\arcmin$ (24~kpc). ``Sim \#1-3''
refer to simulations of models obtained from the spatially resolved
analysis in radial bins 1-3 following our discussion in \S
\ref{bigap}.}
\end{center}
\end{table*}

Thus far we have fitted models to the \xmm\ and \chandra\ spectra of a
particular radial bin independently of the spectra in other radial
bins.  (Although a deprojected model for a specific radial bin does
account for the projected emission from exterior shells, the spectral
fitting is not performed simultaneously with other shells.) Here we
consider models fitted to the total \xmm\ and \chandra\ spectral data
accumulated within radial bins 1-3; i.e., within a circular aperture
of radius, $R=2.5\arcmin$ (24~kpc). We focus on this central region
where there is both \chandra\ and \xmm\ data and evidence for
multiphase gas.  Our objective is to show that the preference for the
multiphase models indicated by the fits to the individual radial bins
1-3 is rendered even more significant when the data in these bins are
analyzed simultaneously in a single aperture.

First we summarize the results (Table \ref{tab.bigap}) of fitting 1T,
2T, and PLDEM models (all 2D for this exercise) to the spectral data
within $R=2.5\arcmin$.  As expected, the 1T model is an even worse fit
than within the individual radial bins 1-3. The best-fitting 1T model
shown in Figure \ref{fig.bigap} displays more pronounced residuals near
1~keV (i.e., characteristic of the ``Fe Bias'') than observed for the
fit only to radial bin 2 shown in Figure \ref{fig.spec.1t}. The
\chandra\ ACIS-S3 data have the most pronounced $\chi^2$ residuals in
Figure \ref{fig.bigap} near 1~keV, but we note that the fractional
residuals between the model and data are very similar for both the
EPIC and ACIS CCDs. Allowing for variable \nh\ does not improve the 1T
fits significantly, despite the fact that a large excess value for the
column density is indicated.

However, the addition of a second temperature component (i.e., 2T
model) provides a vastly improved fit with small residuals near 1~keV
similar to the residuals at other energies (Figure
\ref{fig.bigap}). The PLDEM model fits nearly as well as the 2T model,
and the quality of both of these fits is about as good as could be
hoped for considering the simplicity of these models and that we know
the spectral properties do vary with radius within the $R=2.5\arcmin$
aperture. (Allowing for variable \nh\ also provides negligible
improvement for the multitemperature models even though a large fitted
value of excess \nh\ is indicated -- the 2T result is shown in Table
\ref{tab.bigap}.) 

Now we wish to compare these results to those that should have been
obtained if the real data were actually described by the radially
varying 1T and 2T models in shells 1-3 obtained in \S \ref{1t} and \S
\ref{2t}. To perform this comparison we follow the general approach
discussed in \S 5.3 of \citet{buot99a}. We take the 1T (3D) models in
shells 1-3 obtained from the real EPIC and ACIS data (\S \ref{1t}) and
simulate EPIC and ACIS spectra appropriate for the NGC 5044
observations using the {\sc fakeit} routine in \xspec. These simulated
spectra also contain the projected emission from external shells
obtained from our deprojection analysis. The resulting simulated
source and background pha files for radial bins 1-3 are then summed
(separately for each detector) and analyzed in the same way as done in
Table \ref{tab.bigap}. We refer to this simulated radially varying 1T
model within $R=2.5\arcmin$ as ``Sim \#1'' in Table
\ref{tab.bigap.sims}. ``Sim \#2'' and ``Sim \#3'' in Table
\ref{tab.bigap.sims} are prepared in the same way as ``Sim \#1''
except they refer respectively to the 1T (3D) model with variable \nh\
(\S \ref{nh}) and 2T (3D) model with Galactic \nh\ (\S \ref{2t}).

Fits to Sim \#1 using 1T models give $\chi^2\approx 1800$ which are
much smaller than the values of $\chi^2\approx 4000$ obtained for the
real data in Table \ref{tab.bigap}. The 2T model is also a somewhat
better fit than for the real data and provides a formally acceptable
fit. Although similar to the real data the 2T fit is superior to the
1T fit, and there is negligible improvement when allowing for variable
absorption, the much smaller values of $\chi^2$ for the 1T models
fitted to Sim \#1 rule out the radially varying 1T models with
Galactic absorption. 

Qualitatively different results are obtained for fits to Sim
\#2. Similar to Sim \#1 the fitting the 1T model with Galactic \nh has
yields a much lower value of $\chi^2$ than obtained for the real
data. However, the 1T model with variable \nh\ provides a very large
improvement in the fit -- almost as good as the 2T model with Galactic
\nh. This behavior is totally inconsistent with the results obtained
from the real data in Table \ref{tab.bigap}, and we conclude that the
radially varying 1T model with variable \nh\ is not a viable
description of the spectral data within $R=2.5\arcmin$. 

In contrast, the $\chi^2$ values obtained for fits to Sim \#3 are very
similar to those obtained for the real data in Table
\ref{tab.bigap}. Since only fits to Sim \#3 can reproduce the results
of fitting the real data, this is strong evidence that the
multitemperature models (in this case the radially varying 2T model,
but this also applies to the PLDEM model) are required within the
central $R=2.5\arcmin$ (24~kpc) of NGC 5044.

\section{Discussion: Two-Phase Model}
\label{disc}

\begin{figure*}[t]
\centerline{\psfig{figure=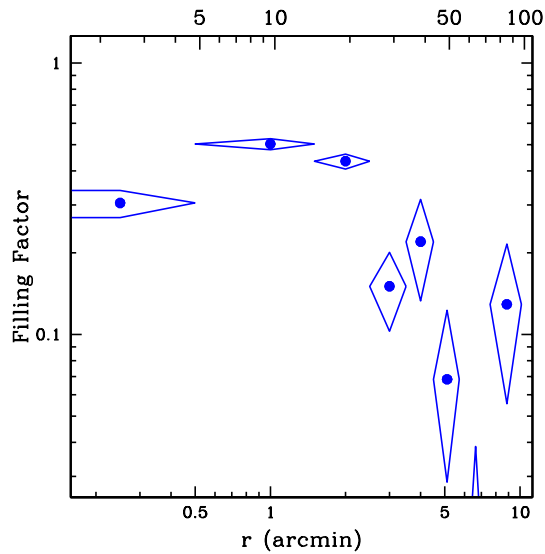,height=0.3\textheight}}
\caption{\label{fig.fillfact}  Volume filling
factor of the cooler component required to maintain pressure
equilibrium between the cooler and hotter phases in the 2T (3D) model
obtained from simultaneous fitting of the \xmm\ and
\chandra\ data. Due to the expected contribution of emission from
discrete sources, the value in the central bin is quite uncertain and
is consistent with unity. The units are arcminutes on the bottom axis
and kpc on the top axis.}
\end{figure*} 

Our spectral deprojection analysis of the \xmm\ and \chandra\ data of
NGC 5044 indicates that a single-phase description of the hot gas is
inadequate for $r\la 30$~kpc. Of the simple multiphase models we
considered, the 2T and PLDEM models provide the best fits in the
central regions. Each of these models describes a limited multiphase
plasma where the cooler temperature components ($\sim 0.7$~keV)
dominate for $r\la 10$~kpc, the contributions of cooler and hotter
($\sim 1.4$~keV) components are similar for $r\approx 20-30$~kpc,
while at larger radii the hotter components dominate so that the gas
is consistent with a single-phase medium.

These results for the 2T model are very similar to those obtained from
an analysis of the \xmm\ data of the group NGC 1399 \citep{buot02a}.
The temperature of the extended hotter component, $\thot\sim 1.4$~keV,
is consistent with the virial temperature of a surrounding group of
mass $\approx 10^{13}\msun$, whereas the temperature of the centrally
concentrated cooler component, $\tcool \sim 0.7$~keV, is similar to
the kinetic temperature of the stars. These parameters suggest a
physical association of the hotter component with the ambient group
gas and the cooler component with stellar ejecta from the dominant
central galaxy.

A potential problem with this scenario is that it might be expected
that these phases should mix very rapidly for a system in
equilibrium. However, the sharp edge in the surface brightness and
isophote center offset for $R\approx 5\arcmin-6\arcmin$ discussed in
\S \ref{image} are suggestive of a ``cold front'' such as has been observed
in several clusters with \chandra\ \citep[e.g.][]{mark02a}. The
surface brightness edge in the MOS image of NGC 5044 near $R\approx
50$~kpc to the NW essentially divides the regions where the cooler and
hotter gas phases are most prominent.

Although the cold front can explain the radial transition from cooler
to hotter dominance, the phases clearly co-exist over a large range in
radius. For this to occur we expect the phases to be in pressure
equilibrium. We have calculated the volume filling factor ($f_{\rm
c}$) of the cooler component of the 2T model required to maintain
pressure equilibrium between the two phases. The result is plotted in
Figure \ref{fig.fillfact}.

For $r\ga 30$~kpc we have $f_{\rm c}\sim 0.1$ indicating that the
cooler gas occupies only a small fraction of the shell volume. In
contrast, for shells 2-3 ($r\sim 5-25$~kpc) we have $f_{\rm c}\approx
0.5$ meaning that each phase occupies half the volume. In shell 1
($r\la 5$~kpc) the value shown ($f_{\rm c}\approx 0.3$) is actually a
lower limit because the value of \thot\ is overestimated due to
contamination from discrete sources (\S \ref{2t}).

\section{Conclusions}
\label{conc}

The spectral deprojection analysis of \xmm\ and \chandra\ data favors
a two-phase (2T) or limited multi-phase medium (PLDEM) within the
central $r \sim 30$~kpc of NGC 5044.  The cooler component in the 2T
models has a temperature $\tcool \sim 0.7$~keV similar to the kinetic
temperature of the stars in the central galaxy NGC 5044, and the
hotter component has a temperature $\thot \sim 1.4$~keV characteristic
of the massive $\sim 10^{13}$~$M_{\odot}$ dark halo of the surrounding
galaxy group.  Nevertheless, both temperature components appear at
every radius $\la 30$~kpc.  In spite of the similarity of the hot
phase temperature \thot\ and the group virial temperature at all
radii, it is likely that gas at small radii with temperature \thot\ is
heated by a central AGN.  Some additional heating at large radii could
arise from the energy associated with establishing the cold front near
$r \sim 30$ kpc, as suggested by the sharp edge visible in the EPIC
MOS and pn images.

The need for two discrete temperatures in the 2T models cannot be
attributed solely to the Fe-L lines at $\sim 1$~keV, which still may
be uncertain in the plasma codes, but is also required by the spectrum
below 0.7 keV. Our comparison of results using the \apec\ and \mekal\
codes in \S \ref{plasma} indicates that even large differences in the
plasma codes have a small effect at the moderate resolution of the
EPIC and ACIS CCDs.

As an alternative, and possibly more physically plausible model, we
have shown that a continuous, but limited, range of temperatures in
each spherical shell (PLDEM) can explain the NGC 5044 data for $r \la
30$~kpc as well as the 2T model. However, the range of temperatures
required in each spherical shell exceeds the radial temperature
variation of the best-fitting single-phase models across the same
shell.  For either type of thermal model there is no evidence at any
radius for gas at temperatures $\la 0.7$~keV. These results are very
similar to our previous 2T models of NGC 5044 using
\asca\ data \citep{buot99a}.  However, within $r \sim 30$~kpc the \asca\
data lacked the spatial resolution of \xmm\ and \chandra\ and could
not distinguish between a single-phase medium in which the gas
temperature varies with radius, a two-phase (2T) medium, or
multi-phase gas at every radius.

The remarkable irregularities visible for $r \la 10$~kpc in the
\chandra\ image of NGC 5044 (Figure \ref{fig.images}) support
the notion of fluctuations in the gas density and, by implication,
also in the gas temperature.  In pressure equilibrium the ratio of
X-ray emissivities in the two phases should be approximately,
$[(n_{\rm e})_{\rm c}/(n_{\rm e})_{\rm h}]^2 \propto (\thot/\tcool)^2
\approx 5$, which may be sufficient to account for the conspicuous
surface brightness fluctuations in Figure \ref{fig.images}.  Within
the central annulus ($R=0.5\arcmin\approx 5$ kpc), which includes
nearly half the optical image of the NGC 5044 galaxy, the X-ray
spectrum is further complicated by hard $\sim 10$~keV bremsstrahlung
radiation from X-ray binary stars, so the precise range of gas
temperatures is less accurately determined in this region.

The 2T or limited multi-phase thermal properties of NGC 5044 are very
similar to those of the bright galaxy group NGC 1399 revealed by \xmm\
\citep{buot02a}.  A 2T model is also preferred by \citet{mole02a} in
his \xmm\ analysis of M87.  However, Molendi argues that the
temperature range $\thot - \tcool$ implied for each annulus by his 2T
model for M87 is consistent with a radially varying single-phase
temperature except in regions where the X-ray image is clearly
distorted by interaction with the central radio source.  Nevertheless,
the pattern that is emerging from observations of NGC 5044 and NGC
1399 may differ substantially from the gasdynamical models of groups
and clusters constructed by \citet{brig99b,brig02a}. In the
gasdynamical models relatively cool ejecta from stellar mass loss mix
thermally on small scales with hotter ambient group or cluster gas to
reproduce the single-phase radial temperature profiles typically
observed in groups and clusters.  Instead, our observations suggest a
limited range of thermal phases that are incompletely mixed at every
radius, which, if viewed as a single-phase gas, can reproduce the same
typical ``cooling flow'' thermal profile but with a reduction in the
quality of the spectral fit.

\acknowledgements It is a pleasure to thank T. Fang for communicating
results of his preliminary analysis of the RGS data of NGC 5044.  We
gratefully acknowledge partial support from NASA grants NAG5-9956,
NAG5-10758, and NAG5-10748.

\bibliographystyle{apj}

\end{document}